\documentclass[journal]{IEEEtran}
\usepackage{amsmath,amsfonts}
\usepackage{algorithm}
\usepackage{array}
\usepackage[caption=false,font=normalsize,labelfont=sf,textfont=sf]{subfig}
\usepackage{textcomp}
\usepackage{stfloats}
\usepackage{url}
\usepackage{verbatim}
\usepackage{graphicx}
\usepackage{cite}
\usepackage{amssymb}   
\usepackage{booktabs}  
\usepackage{multirow}
\usepackage{algpseudocode}

\usepackage{xcolor}
\usepackage[
    colorlinks=true,
    linkcolor=blue,
    citecolor=blue,
    urlcolor=blue,
    breaklinks=true
]{hyperref}



\makeatletter
\renewcommand{\eqref}[1]{\hyperref[#1]{\textup{\color{blue}(\ref*{#1})}}}
\makeatother

\begin{document}

 \title{From Ideal Motion to Flight-Executable Communications: LLM-Evolved Multi-UAV Deployment for Cell-Free Massive MIMO}

\author{
Yuyao~Wang,
Gaoze~Mu,
Yongan~Zheng,
Yanzhao~Hou,~\IEEEmembership{Member,~IEEE},
Qingqing~Wu,~\IEEEmembership{Senior Member,~IEEE},
Qimei~Cui,~\IEEEmembership{Senior Member,~IEEE},
Xiaofeng~Tao,~\IEEEmembership{Senior Member,~IEEE}
and Ping~Zhang,~\IEEEmembership{Fellow,~IEEE}
\thanks{Yuyao~Wang, Gaoze~Mu, and Yongan~Zheng are with the National Engineering Research
Center for Mobile Network Technologies, Beijing University of Posts and Telecommunications, Beijing 100876, China (e-mail: yuyaowang@bupt.edu.cn; mugz@bupt.edu.cn; yonganzheng@bupt.edu.cn).}
\thanks{Yanzhao~Hou, Qimei~Cui, and Xiaofeng~Tao are with the National Engineering Research Center for Mobile Network Technologies, Beijing University of Posts and Telecommunications, Beijing 100876, China, and also with the Department of Broadband Communication, Peng Cheng Laboratory, Shenzhen 518055, China (e-mail: houyanzhao@bupt.edu.cn; cuiqimei@bupt.edu.cn; taoxf@bupt.edu.cn).}
\thanks{Qingqing Wu is with the Department of Electronic Engineering, Shanghai Jiao Tong University, 200240, China (e-mail: qingqingwu@sjtu.edu.cn).}
\thanks{Ping~Zhang is with the State Key Laboratory of Networking and Switching Technology, Beijing University of Posts and Telecommunications, Beijing 100876, China(e-mail: pzhang@bupt.edu.cn).}
\thanks{Corresponding author: Yanzhao~Hou.}
\vspace{-1cm}
}

\maketitle

\begin{abstract}

Cell-free massive multiple-input multiple-output (CF-mMIMO) is a promising paradigm for future wireless networks, providing user-centric services and cooperative coverage. By using unmanned aerial vehicles (UAVs) as aerial access points, CF-mMIMO networks can exploit UAV mobility to enhance three-dimensional (3D) coverage and spectral efficiency (SE). However, most existing studies on UAV deployment for communication optimization typically assume that UAVs follow ideal point-mass motion (IM), neglecting flight-control constraints and finite-horizon position errors in real flight execution. Consequently, IM-trained deployment policies may degrade severely or become difficult to execute in practice. Motivated by this, we model each UAV as a six-degree-of-freedom quadrotor rigid body with a cascaded flight controller to capture the impact of flight-control-constrained motion (FM) on communication optimization. Based on this model, we formulate a joint UAV 3D deployment and power allocation problem under FM to maximize the downlink average SE in CF-mMIMO networks. To address this problem, we propose LERE, a large language model (LLM)-enhanced multi-agent reinforcement learning (MARL) framework. In LERE, the LLM evolves hybrid rewards with both global and local components via multi-level feedback. The evolved hybrid rewards guide MARL policy optimization and promote multi-UAV cooperation. Experimental results demonstrate that LERE achieves higher SE than reward-design baselines while substantially reducing UAV position errors. Notably, when tested under FM execution, the FM-trained LERE policy achieves a 60.49\% SE gain over its IM-trained counterpart, confirming the necessity of incorporating flight-control constraints into UAV-enabled CF-mMIMO optimization.

\end{abstract}

\begin{IEEEkeywords}
UAV, cell-free massive MIMO, large language model, multi-agent reinforcement learning.
\end{IEEEkeywords}

\vspace{-0.15cm}
\section{Introduction}

\IEEEPARstart{C}{ell-free} massive multiple-input multiple-output (CF-mMIMO) networks have recently emerged as a promising architecture for future ubiquitous wireless access for massive Internet of Things (IoT) applications~\cite{elhoushy2021cell}. In contrast to conventional cell-centric networks, CF-mMIMO coordinates geographically distributed access points (APs) through a central processing unit (CPU), allowing multiple APs to cooperatively serve user equipment (UE) over the same time-frequency resources~\cite{shah2026joint,wan2024performance}. This architecture can effectively mitigate severe inter-cell interference and frequent handovers, which are common in conventional cellular networks~\cite{shi2022meta,wang2022deployment}.

By integrating unmanned aerial vehicles (UAVs) as aerial APs, CF-mMIMO networks can further exploit UAV mobility to enhance flexible three-dimensional (3D) coverage, especially in hotspot, emergency, low-altitude, and infrastructure-limited scenarios~\cite{shah2026joint,wan2024performance}. Different from fixed terrestrial APs, UAVs can dynamically adjust their spatial positions according to UE distributions and channel conditions, thereby improving line-of-sight (LoS) connectivity and spectral efficiency (SE). Therefore, UAV-enabled CF-mMIMO networks provide a highly flexible and cooperative networking paradigm for future wireless systems.

The performance gain of UAV-enabled CF-mMIMO networks largely depends on how UAVs are deployed and how transmit power is allocated. On the one hand, UAV 3D deployment directly affects large-scale channel gains, LoS probabilities, and the spatial relationship between UAV APs and ground UEs~\cite{khawaja2019survey}. On the other hand, downlink (DL) power allocation determines the service strength of different UEs and regulates inter-user interference under cooperative transmission. As a result, joint UAV 3D deployment and DL power allocation is essential for maximizing the average SE in UAV-assisted CF-mMIMO networks. Motivated by this, existing studies have investigated UAV deployment, trajectory design, and communication resource allocation in various scenarios, including data collection, mobile edge computing (MEC), integrated sensing and communication (ISAC) in CF-mMIMO networks~\cite{chen2022joint,lv2025large,gao2024marl,xu2023soft,shah2026joint}.

However, most existing studies on UAV deployment and communication optimization rely on ideal point-mass motion (IM) assumptions, where UAVs are simplified as point masses that can accurately reach the target positions generated by communication optimization algorithms within each discrete time slot~\cite{chen2022joint,lv2025large,gao2024marl,2025trajectory,shah2026joint,xu2023soft}. Such an idealized motion model neglects the practical flight-control constraints of real UAV platforms. In practical low-altitude applications, UAVs are commonly implemented as quadrotors due to their hovering and maneuvering capabilities. The motion of a quadrotor is governed by six-degree-of-freedom (6-DoF) rigid-body dynamics and flight-control algorithms~\cite{luukkonen2011modelling,romero2022time}. Consequently, a quadrotor may not exactly reach the commanded target position within a finite execution horizon, and the resulting position error depends on both the current flight state and the target command.

This mismatch between ideal deployment commands and flight-control-constrained execution can significantly affect communication performance. In CF-mMIMO networks, even small deviations between target and realized UAV positions may change multiple UAV-UE channel gains, LoS link conditions, and the interference structure. Moreover, the DL power allocation optimized for the target UAV deployment may become mismatched with the realized channels after practical flight execution. Therefore, policies optimized under IM assumptions may suffer from severe SE degradation or even become difficult to execute in real UAV systems. Despite its practical importance, UAV communication optimization under flight-control-constrained motion remains insufficiently investigated.

To address this issue, we model each UAV as a 6-DoF quadrotor rigid body equipped with a cascaded flight controller. This practical model captures the finite-horizon execution behavior of UAVs and characterizes the position errors induced by flight-control-constrained motion, referred to as FM in this paper. Based on this model, we formulate a flight-control-constrained joint optimization problem of multi-UAV 3D deployment and DL power allocation to maximize the DL average SE in CF-mMIMO networks. In this problem, deployment commands determine the finite-horizon flight execution of quadrotors, realized UAV positions determine the actual channel conditions, and DL power allocation regulates the cooperative transmission and inter-user interference. Therefore, communication performance, flight executability, and multi-UAV coordination are tightly coupled, making the formulated problem highly non-convex and challenging to solve.

Classical optimization methods, such as successive convex approximation (SCA)~\cite{wu2018joint}, usually require explicit mathematical transformations and convex approximations. However, such transformations are difficult to design for the considered problem because of the coupled continuous variables, nonlinear channel variations, practical flight-control dynamics, and multi-UAV interactions. Reinforcement learning (RL) provides a data-driven alternative for learning continuous control policies through interactions with the environment, especially when explicit convex reformulation is intractable~\cite{wang2025drl}. Once trained, an RL policy can generate online decisions through fast neural network inference. For multi-UAV communication optimization, multi-agent reinforcement learning (MARL) is particularly suitable, as it decomposes the large-scale joint action space into agent-wise policies while enabling coordination among UAVs through centralized training and decentralized execution~\cite{li2022applications,zhong2021multi}.

Nevertheless, the effectiveness of RL methods, especially MARL, hinges critically on the quality of reward design, which provides the primary guidance signal for policy optimization~\cite{eschmann2021reward}. For the considered flight-control-constrained CF-mMIMO optimization problem, the reward should not only reflect the global system-level SE objective, but also provide informative local feedback for each UAV regarding its contribution, flight execution behavior, and constraint violations. A purely global reward is consistent with the network-wide objective but may provide weak and delayed feedback to individual agents. In contrast, purely local rewards can reflect agent-specific behaviors but may neglect the global communication objective and aggravate the credit-assignment problem~\cite{du2019liir}. Therefore, an effective hybrid reward mechanism with both global and local components is required. However, manually designing such rewards is a tedious trial-and-error process that requires substantial domain expertise and may lead to ineffective guidance or unexpected behaviors~\cite{hadfield2017inverse,booth2023perils}.

Recently, large language models (LLMs) have shown great potential for enhancing RL in complex wireless communication scenarios, where conventional RL methods often suffer from inefficient exploration, limited generalization, and difficult reward design. By leveraging pretrained knowledge, contextual reasoning, and code-generation capabilities, LLMs can generate task-specific reward functions and provide informative guidance for policy optimization~\cite{cai2025tutorial,zheng2026large,cai2025large,cui2025overview}. Existing LLM-driven reward-design studies mainly focus on general RL tasks and still face several limitations when applied to MARL-based wireless network optimization~\cite{kwon2023reward,xie2024text2reward,ma2024eureka,li2025efficient}. For example, some methods use LLMs as online proxy reward models, which incurs high inference cost and latency~\cite{kwon2023reward}. Some methods depend on human feedback and therefore cannot achieve fully automated reward evolution~\cite{xie2024text2reward}. Other methods generate multiple reward candidates through heuristic evolutionary search and require separate policy training for each candidate, resulting in high training overhead and limited feedback efficiency~\cite{ma2024eureka,li2025efficient}. These limitations motivate the design of an efficient LLM-enhanced MARL framework tailored to hybrid reward evolution in multi-UAV CF-mMIMO optimization.

In this paper, we propose LERE, an LLM-enhanced MARL framework for flight-control-constrained joint UAV deployment and power allocation in CF-mMIMO networks. LERE employs the LLM to evolve hybrid rewards with both global and local components through multi-level feedback, thereby providing more informative guidance for MARL policy optimization and promoting cooperative decision-making among multiple UAVs. The main contributions of this paper are summarized as follows.

\begin{itemize}
    \item First, we incorporate practical flight-control constraints into UAV-enabled CF-mMIMO network optimization by modeling each UAV as a 6-DoF quadrotor rigid body with a cascaded flight controller. This enables the formulated problem to capture finite-horizon position errors caused by practical flight execution, rather than relying on ideal point-mass motion assumptions.

    \item Second, we formulate a joint UAV 3D deployment and DL power allocation problem under FM for maximizing the downlink average SE. The formulation explicitly couples target deployment commands, realized UAV positions, channel conditions, and power allocation, thereby reflecting the interaction between flight executability and communication performance.

    \item Third, we propose LERE, an LLM-enhanced MARL framework that evolves hybrid rewards through multi-level feedback. The designed hybrid reward mechanism effectively balances global SE-oriented objectives and local UAV-specific incentives, improving policy learning efficiency and enhancing multi-UAV cooperation.
   
    \item Finally, experimental results demonstrate that LERE achieves higher SE than reward-design baselines while substantially reducing UAV position errors. In particular, when evaluated under flight-control-constrained execution, the FM-trained LERE policy achieves a 60.49\% SE gain over its IM-trained counterpart, confirming the necessity of incorporating practical flight-control constraints into UAV-enabled CF-mMIMO optimization.
\end{itemize}

\section{System Model}
\label{system_model}

\begin{figure}[!t]
  \centering
  \includegraphics[width=8.5cm]{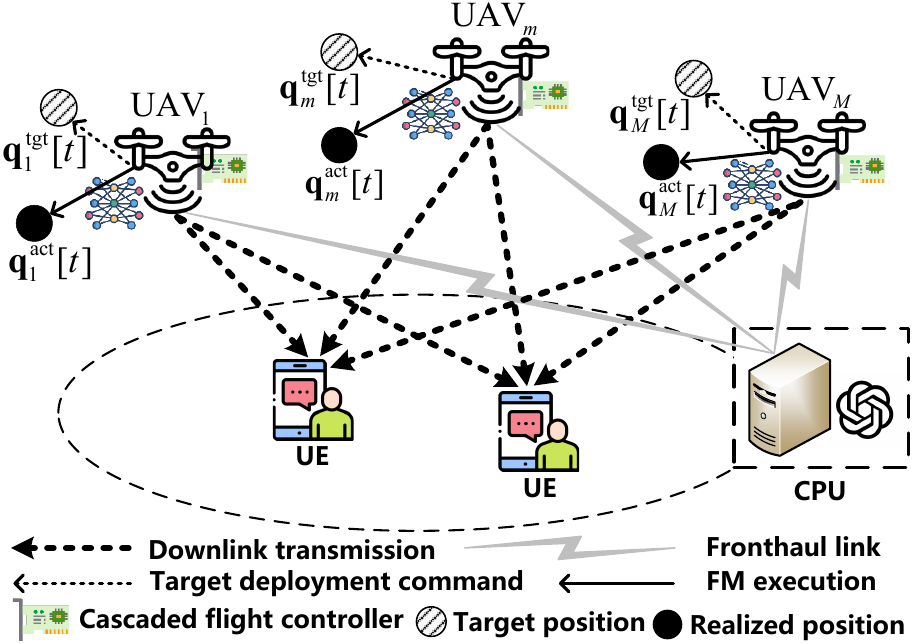}
  \caption{Flight-control-constrained joint UAV 3D deployment and power allocation system model.}
  \label{system}
\end{figure}

As shown in Fig.~\ref{system}, we consider the DL transmission of a UAV-enabled CF-mMIMO network in a 3D region, where $M$ UAVs, indexed by $m\in\mathcal{M}=\{1,\ldots,M\}$, act as aerial APs and cooperatively serve $K$ single-antenna ground UEs, indexed by $k\in\mathcal{K}=\{1,\ldots,K\}$. The UAV APs are connected to the CPU through dedicated high-capacity fronthaul links, following common UAV-enabled CF network architectures~\cite{wan2024performance,shah2026joint,xu2023soft}. These links are modeled as reliable and capacity-sufficient, and they are orthogonal to the access links between UAV APs and UEs~\cite{wan2024performance}. The ground service area is denoted as
$\mathcal{A}=[0~{\rm m},1000~{\rm m}]\times[0~{\rm m},1000~{\rm m}]$, and the feasible UAV deployment region is 
$\mathcal{Q}=[0~{\rm m},1000~{\rm m}]\times[0~{\rm m},1000~{\rm m}]\times[50~{\rm m},200~{\rm m}]$. Each UAV AP is equipped with an $N$-antenna uniform linear array (ULA). The position of UE $k$ is denoted by $\mathbf{u}_{k}=[x_k,y_k,1.5]^T~{\rm m}$, where $(x_k,y_k)\in\mathcal{A}$.

The system operates in discrete time slots indexed by $t$. At the beginning of slot $t$, UAV $m$ has the quadrotor state $\mathbf{x}_m[t]$, and the deployment policy assigns a commanded target position $\mathbf{q}_{m}^{\rm tgt}[t]\in\mathcal{Q}$. The interval from slot $t$ to slot $t+1$ is a finite execution horizon $T_{\rm f}=N_{\rm f}h$, where $h$ is the flight-control step size, $N_{\rm f}$ is the number of flight-control iterations, and $r\in\{0,\ldots, N_{\rm f}-1\}$ indexes the control step. 

During this execution horizon, the quadrotor flight-control process described in Section~\ref{sec:flight_control} updates the UAV state from $\mathbf{x}_m[t]$ to $\mathbf{x}_m[t+1]$ over $N_{\rm f}$ closed-loop iterations and yields the realized position $\mathbf{q}_{m}^{\rm act}[t]\in\mathcal{Q}$. This differs from the IM model, which idealizes the execution process by assuming 
$\mathbf{q}_{m}^{\rm act}[t]=\mathbf{q}_{m}^{\rm tgt}[t]$. Accordingly, all air-to-ground (A2G) channels, precoders, and DL SE values are evaluated using the realized UAV positions 
$\mathbf{Q}^{\rm act}[t]=\{\mathbf{q}_{m}^{\rm act}[t]\}_{m=1}^{M}$ 
rather than the commanded target positions 
$\mathbf{Q}^{\rm tgt}[t]=\{\mathbf{q}_{m}^{\rm tgt}[t]\}_{m=1}^{M}$.

The considered system operates in time-division duplex (TDD) mode. Each coherence block contains $\tau_c$ channel uses, where $\tau_p$ channel uses are used for uplink (UL) pilot transmission and minimum mean-square error (MMSE) channel estimation. The remaining $\tau_c-\tau_p$ channel uses are used for DL data transmission with local MMSE (L-MMSE) precoding and power allocation. For notational simplicity, the time index $t$ is omitted in the following subsections.

\vspace{-0.3cm}
\subsection{Channel Model}
\label{channel_model}

The A2G Rician fading channel between UAV AP $m$ and UE $k$ is modeled as $\mathbf{g}_{mk} = \bar{\mathbf{g}}_{mk} + \tilde{\mathbf{g}}_{mk} \in \mathbb{C}^{N}$ \cite{wang2024optimal}. 
The deterministic LoS component is given by 
\begin{equation}
\bar{\mathbf{g}}_{mk} = \sqrt{\frac{\kappa_{mk}}{\kappa_{mk}+1}\beta_{mk}}\, \mathbf{a}(\varphi_{mk}) \in \mathbb{C}^{N}, 
\label{eq:los}
\end{equation}
where $\kappa_{mk}$ is the Rician factor and $\beta_{mk}$ represents the large-scale fading gain. $\mathbf{a}(\varphi_{mk}) = \left[ e^{j2\pi(i-1)\frac{d_a}{\lambda}\sin\varphi_{mk}} \right]_{i=1}^{N} \in \mathbb{C}^{N}$ denotes the array steering vector, where $\varphi_{mk}$ is the azimuth angle from UAV AP $m$ to UE $k$, $d_a$ is the antenna element spacing, and $\lambda$ is the carrier wavelength.

The stochastic NLoS component is modeled as
\begin{equation}
\tilde{\mathbf{g}}_{mk}=\sqrt{\frac{\beta_{mk}}{1+\kappa_{mk}}}\mathbf{h}_{mk}^{\mathrm{NLoS}},\quad \mathbf{h}_{mk}^{\mathrm{NLoS}}\sim\mathcal{CN}\left(\mathbf{0},\mathbf{R}_{mk}^{\mathrm{NLoS}}\right),
\label{eq:nlos}
\end{equation}
where $\mathbf{R}_{mk}^{\mathrm{NLoS}}$ is the spatial correlation matrix generated by the local scattering model. Hence, $\tilde{\mathbf{g}}_{mk}\sim\mathcal{CN}(\mathbf{0},\tilde{\mathbf{R}}_{mk})$ with $\tilde{\mathbf{R}}_{mk}=\frac{\beta_{mk}}{1+\kappa_{mk}}\mathbf{R}_{mk}^{\mathrm{NLoS}}$.  

Let $d_{mk}=\|\mathbf{q}_{m}^{\rm act}-\mathbf{u}_k\|_2$ denote the distance between UAV $m$ and UE $k$. The elevation angle is $\alpha_{mk}$, and the LoS probability is modeled as $P_{\mathrm{LoS},mk}=\frac{1}{1+\vartheta\exp\!\left[-\zeta\left(\frac{180}{\pi}\alpha_{mk}-\vartheta\right)\right]}$, where $\vartheta$ and $\zeta$ are environment-dependent parameters, and the Rician factor is given by $\kappa_{mk}=\min\!\left\{10^{1.3-0.003d_{mk}},\frac{P_{{\rm LoS},mk}}{1-P_{{\rm LoS},mk}}\right\}$ \cite{2025trajectory}. The large-scale fading gain between UAV AP $m$ and UE $k$ is modeled in dB as \cite{2025trajectory} 
\begin{equation}
\begin{aligned}
\beta_{mk}^{\rm dB} &= -20 \log_{10}\!\left(\frac{4\pi f_c d_{mk}}{c_0}\right) \\
&\quad - P_{{\rm LoS},mk}\eta_{\rm LoS} 
 - \left(1-P_{{\rm LoS},mk}\right)\eta_{\rm NLoS},
\end{aligned}
\label{eq:beta}
\end{equation}
where $\beta_{mk}=10^{\beta_{mk}^{\rm dB}/10}$, $f_c$ is the carrier frequency, $c_0$ is the speed of light, and $\eta_{\rm LoS}$ and $\eta_{\rm NLoS}$ denote the excess losses for LoS and NLoS propagation, respectively.

\vspace{-0.4cm}
\subsection{Uplink Channel Estimation}
\label{channel_estimation}
Since the considered system operates in TDD mode, we first analyze uplink (UL) channel estimation, through which the DL channel can be obtained by exploiting the channel reciprocity. 

For UL channel estimation, we use $\tau_p$ mutually orthogonal pilot sequences 
$\{\boldsymbol{\phi}_1,\ldots,\boldsymbol{\phi}_{\tau_p}\}$, where 
$\boldsymbol{\phi}_w\in\mathbb{C}^{\tau_p}$ and $\|\boldsymbol{\phi}_w\|^2=\tau_p$.
Since $\tau_p<K$, multiple UEs may share the same pilot. 
Let $w_k\in\{1,\ldots,\tau_p\}$ be the pilot index of UE $k$, and define $\mathcal{P}_k=\{l:w_l=w_k\}$ as the set of UEs sharing this pilot. The pilot allocation strategy follows from that of~\cite{ozdogan2019performance}. All UEs send their pilot signals simultaneously. After despreading at AP $m$, the processed observation vector for UE $k$ is given by \cite{wang2024optimal} $\mathbf{y}_{p,mk} = \sum_{l \in \mathcal{P}_k} \tau_p \sqrt{p_l^{\rm ul}} \mathbf{g}_{ml} + \mathbf{n}_{p,mk} \in \mathbb{C}^{N}$, where $p_l^{\rm ul}$ represents the UL transmit power of UE $l$, and $\mathbf{n}_{p,mk} \sim \mathcal{CN}(\mathbf{0}, \tau_p \sigma^2 \mathbf{I}_N)$ is the thermal noise vector with $\sigma^2$ being the noise power. The MMSE estimate of $\mathbf g_{mk}$ is~\cite{wang2024optimal}
\begin{equation}
\hat{\mathbf{g}}_{mk} = \bar{\mathbf{g}}_{mk} + \sqrt{p_k^{\rm ul}} \tilde{\mathbf{R}}_{mk} \mathbf{\Psi}_{mk}^{-1} (\mathbf{y}_{p,mk} - \bar{\mathbf{y}}_{p,mk}),
\label{eq:mmse_estimator}
\end{equation}
where $\bar{\mathbf{y}}_{p,mk} = \sum_{l \in \mathcal{P}_k} \sqrt{p_l^{\rm ul}} \tau_p \bar{\mathbf{g}}_{ml}$, $\mathbf{\Psi}_{mk} = \frac{1}{\tau_p} \mathbb{E}\{ (\mathbf{y}_{p,mk} - \bar{\mathbf{y}}_{p,mk}) (\mathbf{y}_{p,mk} - \bar{\mathbf{y}}_{p,mk})^H \} = \sum_{l \in \mathcal{P}_k} \tau_p p_l^{\rm ul} \tilde{\mathbf{R}}_{ml} + \sigma^2 \mathbf{I}_N$. Let $\mathbf{e}_{mk} = \mathbf{g}_{mk} - \hat{\mathbf{g}}_{mk}$ denote the estimation error vector. $\mathrm{Cov}\{\mathbf{e}_{mk}\} = \mathbf{C}_{mk} = \tilde{\mathbf{R}}_{mk} - p_k^{\rm ul} \tau_p \tilde{\mathbf{R}}_{mk} \mathbf{\Psi}_{mk}^{-1} \tilde{\mathbf{R}}_{mk}$ \cite{wang2024optimal, wang2025optimal}.
\vspace{-0.6cm}
\subsection{Downlink Data Transmission}
\label{downlink}
In the DL data transmission phase, AP $m$ transmits $\mathbf{s}_m = \sum_{k=1}^{K} \mathbf{f}_{mk} \varsigma_k = \sum_{k=1}^{K} \eta_{mk} \mathbf{v}_{mk} \varsigma_k$. Here, $\varsigma_k \sim \mathcal{CN}(0, 1)$ denotes the intended data symbol for UE $k$, and $\mathbf{f}_{mk} \in \mathbb{C}^N$ is the power-scaled precoding vector for UE $k$. The DL precoding vector from AP $m$ to UE $k$ is denoted by $\mathbf{v}_{mk}$, while $\eta_{mk} = \sqrt{\frac{p_{mk}^{{\rm dl}}}{\mathbb{E}\{ \| \mathbf{v}_{mk} \|^2 \}}}$ is the power normalization coefficient. $p_{mk}^{\rm dl}$ is the DL transmit power allocated by AP $m$ to UE $k$. The transmit power budget at AP $m$ is $p_m$, with $\sum_{k=1}^K \mathbb{E}\{\|\mathbf{f}_{mk}\|^2\} \le p_m$.

Leveraging channel reciprocity in TDD, $\mathbf{v}_{mk}$ can be constructed from the UL L-MMSE combining formulation as \cite{wang2025optimal}
\begin{equation}
\resizebox{0.88\linewidth}{!}{$\displaystyle
\mathbf{v}_{mk}
=p_k^{\rm ul}\!\left(
\sum_{l=1}^{K}p_l^{\rm ul}
(\hat{\mathbf{g}}_{ml}\hat{\mathbf{g}}_{ml}^{H}+\mathbf{C}_{ml})
+\sigma^2\mathbf{I}_N
\right)^{-1}\!\hat{\mathbf{g}}_{mk}.
$}
\label{eq:lmmse}
\end{equation}

The aggregate signal received at UE $k$ is given by $y_k = \sum_{m=1}^{M} \mathbf{g}_{mk}^H \mathbf{s}_m + n_k = \sum_{m=1}^{M} \mathbf{g}_{mk}^H \mathbf{f}_{mk} \varsigma_k + \sum_{l \neq k}^{K} \sum_{m=1}^{M} \mathbf{g}_{mk}^H \mathbf{f}_{ml} \varsigma_l + n_k$. $n_k \sim \mathcal{CN}(0, \sigma^2)$ is the DL noise of UE $k$. By invoking the use-and-then-forget (UatF) bounding technique \cite{bjornson2017massive}, the achievable DL SE for UE $k$ is given by \cite{wang2024optimal} $S_k^{\mathrm{dl}} = \frac{\tau_c -\tau_p}{\tau_c} \log_2 \left( 1 + \mathrm{SINR}_k^{\mathrm{dl}} \right)$. Here, the effective signal-to-interference-plus-noise ratio (SINR) is given by \cite{wang2024optimal}
\begin{equation}
\resizebox{0.91\linewidth}{!}{$ 
\displaystyle 
\mathrm{SINR}_k^{\mathrm{dl}} = \frac{\left| \sum_{m=1}^{M} \mathbb{E}\{ \mathbf{f}_{mk}^H \mathbf{g}_{mk} \} \right|^2}{\sum_{l=1}^{K} \mathbb{E}\left\{ \left| \sum_{m=1}^{M} \mathbf{f}_{ml}^H \mathbf{g}_{mk} \right|^2 \right\} - \left| \sum_{m=1}^{M} \mathbb{E}\{ \mathbf{f}_{mk}^H \mathbf{g}_{mk} \} \right|^2 + \sigma^2}.
$}
\label{eq:dl_sinr}
\end{equation}

\section{Quadrotor Flight Control}
\label{sec:flight_control}
As discussed above, existing studies typically assume that UAVs follow the IM model, where commanded target positions are reached exactly within each $T_{\rm f}$. This assumption ignores the finite-horizon response of practical quadrotors governed by rigid-body dynamics and flight-control algorithms. 

In contrast, this section establishes a finite-horizon quadrotor execution model that maps each commanded target position $\mathbf{q}^{\rm tgt}$ to the realized UAV position $\mathbf{q}^{\rm act}$ used for communication evaluation. Specifically, we first introduce the coordinate frames, motor actuation, and 6-DoF rigid-body dynamics, and then present the cascaded flight controller and closed-loop execution process in Algorithm~\ref{alg:fm_execution}. For notational simplicity, the UAV index $m$ and slot index $t$ are omitted in this section.

\vspace{-0.5cm}
\subsection{Quadrotor Dynamics and Actuation Model}
\label{subsec:quadrotor_model}

\subsubsection{Coordinate Frames and State Vector}
\begin{figure}[!t]
  \centering
  \includegraphics[width=0.8\linewidth]{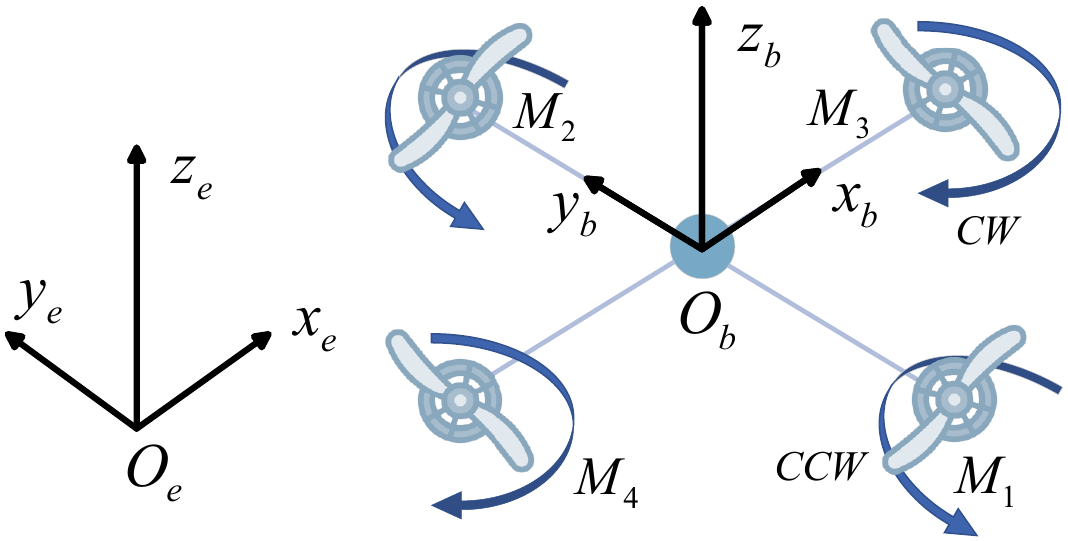}
  \caption{The inertial and body frames of a plus configuration quadrotor.} 
  \label{quadrotor} 
\end{figure}

As shown in Fig.~\ref{quadrotor}, the inertial frame $\mathcal{E}=\{O_e,x_e,y_e,z_e\}$ follows the east-north-up (ENU) convention, where $z_e$ points opposite to gravity. The body frame is $\mathcal{B}=\{O_b,x_b,y_b,z_b\}$. For the plus configuration, $+x_b$ points to motor $M_3$ and $+y_b$ points to motor $M_2$. Motors $M_1$ and $M_2$ rotate counter-clockwise (CCW), while motors $M_3$ and $M_4$ rotate clockwise (CW). The 12D quadrotor state is represented by $\mathbf{x}=[\mathbf{q}^T,\mathbf{v}^T,\boldsymbol{\Theta}^T,\boldsymbol{\omega}^T]^T\in\mathbb{R}^{12}$, where $\mathbf{q},\mathbf{v}\in\mathbb{R}^{3}$ denote the position and velocity in $\mathcal{E}$, respectively. The attitude is represented by Euler angles $\boldsymbol{\Theta}=[\phi,\theta,\psi]^T$ for roll, pitch, and yaw. The angular-rate vector is denoted by $\boldsymbol{\omega}\in\mathbb{R}^{3}$ in the body frame $\mathcal{B}$.

\subsubsection{Motor Actuation Model}
\label{subsec:actuator}
At control step $r$, given the motor command $u_{r,i}\in[0,1]$, the rotor speed $\varpi_{r,i}$ is updated for each rotor $i\in\{1,\ldots,4\}$. The motor response lag is captured by the following first-order rotor-speed dynamics \cite{bouabdallah2007full}
\begin{equation}
    \dot{\varpi}_{r,i} = f_{\varpi}(\varpi_{r,i}, u_{r,i}) = \frac{1}{T_m} \left( C_{\varpi} u_{r,i} + \varpi_b - \varpi_{r,i} \right),
    \label{eq:motor_ode}
\end{equation}
where $\dot{\varpi}_{r,i}$ denotes the rotor angular acceleration, $T_m$ is the motor time constant, $C_{\varpi}$ is the throttle-to-speed gain, and $\varpi_b$ is the rotor-speed bias.
The rotor-speed dynamics are discretized by 4th-order Runge-Kutta (RK4) with step size $h$ \cite{evans1991new}. The rotor speed is updated as
\begin{equation}
\varpi_{r,i}^{'}
=
\operatorname{RK4}\!\left(f_{\varpi},\varpi_{r,i},u_{r,i},h\right).
\label{eq:rk4speed}
\end{equation}

The rotor thrust $T_{r,i}$ and reaction torque $W_{r,i}$ are then computed as $T_{r,i}=C_T(\varpi_{r,i}^{'})^2$ and $W_{r,i}=C_M(\varpi_{r,i}^{'})^2$, where $C_T$ and $C_M$ are the thrust and reaction-torque coefficients, respectively. For the plus configuration, $T_{r,i}$ and $W_{r,i}$ are mapped as~\cite{hoffmann2007quadrotor}
\begin{align}
f_{r,\rm s} &= \sum_{i=1}^{4}T_{r,i},
\tau_{r,\rm x} = L(T_{r,2}-T_{r,1}),
\tau_{r,\rm y} = L(T_{r,4}-T_{r,3}), \notag\\
\tau_{r,\rm z} &= W_{r,4}+W_{r,3}-W_{r,2}-W_{r,1},
\label{eq:rotor_mixing}
\end{align}
where $L$ is the arm length, $f_{r,\rm s}$ is the total thrust, and $\mathbf{n}_{r}=[f_{r,\rm s},(\boldsymbol{\tau}_r)^T]^T$ denotes the generalized control input, with $\boldsymbol{\tau}_r=[\tau_{r,\rm x},\tau_{r,\rm y},\tau_{r,\rm z}]^T$ containing the roll, pitch, and yaw torques.

\begin{algorithm}[t]
\caption{Closed-Loop Flight-Control Execution}
\label{alg:fm_execution}
\footnotesize
\begin{algorithmic}[1]
\Require Initial state $\mathbf{x}_{0}$, target state $\mathbf{x}^{\rm tgt}$, iterations $N_{\rm f}$, and step size $h$.
\Ensure Realized position $\mathbf{q}^{\rm act}$ and final state $\mathbf{x}_{N_{\rm f}}$.

\For{$r=0,\ldots,N_{\rm f}-1$}
    \State Perform position and velocity loop control using Eq.~\eqref{eq:pos_vel_error}.
    \State Generate $[\phi_{r,\rm d},\theta_{r,\rm d}]^T$ and $u_{r,\rm th}$.
    \State Perform attitude and angular-rate loop control using Eq.~\eqref{eq:att_rate_ctrl}.
    \State Mix $u_{r,\rm th}$ and $\mathbf{c}_{r,\omega}$ into motor throttle commands $\mathbf{u}_{r}$ by Eq.~\eqref{eq:allocation}.
    \State Update rotor speeds using Eqs.~\eqref{eq:motor_ode}--\eqref{eq:rk4speed}.
    \State Compute rotor thrusts $T_{r,i}$ and reaction torques $W_{r,i}$.
    \State Map $\{T_{r,i},W_{r,i}\}_{i=1}^{4}$ to $\mathbf{n}_{r}=[f_{r,\rm s},(\boldsymbol{\tau}_{r})^T]^T$ by Eq.~\eqref{eq:rotor_mixing}.
    \State Compute $\dot{\mathbf{x}}_{r}=F(\mathbf{x}_{r},\mathbf{n}_{r})$ 
from Eqs.~\eqref{eq:kinematics}--\eqref{eq:rot_dyn}.
    \State Update the quadrotor state $\mathbf{x}_{r+1}$ by Eq.~\eqref{eq:state_rk4}.
\EndFor

    \State Extract the final position $\mathbf{q}_{N_{\rm f}}$ from $\mathbf{x}_{N_{\rm f}}$, and set $\mathbf{q}^{\rm act}=\mathbf{q}_{N_{\rm f}}$.
\end{algorithmic}
\end{algorithm}

\subsubsection{Quadrotor 6-DoF Rigid-Body Dynamics}
\label{subsec:rigid_body}

At control step $r$, the generalized control input $\mathbf{n}_r$ drives the quadrotor dynamics
$\dot{\mathbf{x}}_r=F(\mathbf{x}_r,\mathbf{n}_r)$. The state derivative is written as
$\dot{\mathbf{x}}_r=[\dot{\mathbf{q}}_r^T,\dot{\mathbf{v}}_r^T,\dot{\boldsymbol{\Theta}}_r^T,\dot{\boldsymbol{\omega}}_r^T]^T$. First, the time derivatives of position and attitude are given by
\begin{equation}
    \dot{\mathbf{q}}_r = \mathbf{v}_r, \quad 
    \dot{\boldsymbol{\Theta}}_r = \mathbf{T}(\boldsymbol{\Theta}_r)\boldsymbol{\omega}_r,
\label{eq:kinematics}
\end{equation}
where $\mathbf{T}(\boldsymbol{\Theta}_r)$ is the body-rate-to-Euler-rate transformation matrix. Governed by the Newton-Euler equations \cite{luukkonen2011modelling, romero2022time}, the translational acceleration $\dot{\mathbf{v}}_r$ in the inertial frame $\mathcal{E}$ is 
\begin{equation}
    \dot{\mathbf{v}}_r = \frac{1}{m_{\rm q}} \mathbf{R}_{\mathcal{B}}^{\mathcal{E}}(\boldsymbol{\Theta}_r) (f_{r,\rm s} \mathbf{e}_3) - g \mathbf{e}_3 + \mathbf{d}_{r, \rm v},
\label{eq:trans_dyn}
\end{equation}
where $m_{\rm q}$ is the mass, $g$ is the gravitational acceleration, $\mathbf{e}_3=[0,0,1]^T$, $\mathbf{R}_{\mathcal{B}}^{\mathcal{E}}(\boldsymbol{\Theta}_r)$ is the body-to-inertial rotation matrix, and $\mathbf{d}_{r, \rm v}$ is the translational disturbance. The rotational angular acceleration $\dot{\boldsymbol{\omega}}_r$ in the body frame $\mathcal{B}$ is formulated as
\begin{equation}
    \dot{\boldsymbol{\omega}}_r = \mathbf{J}^{-1} \left( \boldsymbol{\tau}_r - \boldsymbol{\omega}_r \times (\mathbf{J} \boldsymbol{\omega}_r) - \boldsymbol{\tau}_{r,\rm gyro} + \mathbf{d}_{r,\omega} \right),
\label{eq:rot_dyn}
\end{equation}
where $\mathbf{J}$ is the diagonal inertia tensor, $\mathbf{d}_{r,\omega}$ is the torque disturbance, and 
$\boldsymbol{\tau}_{r,\rm gyro}=J_{\rm in}\varpi_{r,\rm res}(\boldsymbol{\omega}_r\times\mathbf{e}_3)$ is the rotor gyroscopic torque. Here, $J_{\rm in}$ is the rotor inertia and, given the adopted rotation directions, $\varpi_{r,\rm res}=\varpi_{r,4}^{'}+\varpi_{r,3}^{'}-\varpi_{r,2}^{'}-\varpi_{r,1}^{'}$.

Eqs.~\eqref{eq:kinematics}--\eqref{eq:rot_dyn} define the state derivative function $\dot{\mathbf{x}}_r=F(\mathbf{x}_r,\mathbf{n}_r)$ used in the FM simulator. The quadrotor state is advanced by RK4 integration with step size $h$
\begin{equation}
\mathbf{x}_{r+1}
=
\operatorname{RK4}\!\left(F,\mathbf{x}_{r},\mathbf{n}_r,h\right).
\label{eq:state_rk4}
\end{equation}

\vspace{-0.5cm}
\subsection{Cascaded Flight Controller}
\label{subsec:control_strategy}

We employ a cascaded flight controller~\cite{hoffmann2007quadrotor} to drive the quadrotor toward the target state. For the considered deployment task, the controlled target variables are the target position $\mathbf{q}^{\rm tgt}=[x^{\rm tgt},y^{\rm tgt},z^{\rm tgt}]^T$ and the yaw angle 
$\psi^{\rm tgt}=\operatorname{atan2}(y^{\rm tgt}-y_{0},x^{\rm tgt}-x_{0})$, which points from the initial position $\mathbf{q}_0=[x_{0},y_{0},z_{0}]^T$ to the target position in the horizontal plane. The target state is defined as 
$\mathbf{x}^{\rm tgt}=[(\mathbf{q}^{\rm tgt})^{T},\mathbf{0}^{T},[0,0,\psi^{\rm tgt}],\mathbf{0}^{T}]^T$, 
where the target velocity, angular velocity, roll, and pitch are set to zero to represent stable hovering at the target position for communication evaluation.

At control step $r$, the controller takes the current state $\mathbf{x}_{r}=[(\mathbf{q}_{r})^T,(\mathbf{v}_{r})^T,(\boldsymbol{\Theta}_{r})^T,(\boldsymbol{\omega}_{r})^T]^T$, 
where $\mathbf{q}_{r}$, $\mathbf{v}_{r}$, $\boldsymbol{\Theta}_{r}$, and $\boldsymbol{\omega}_{r}$ denote the current position, linear velocity, attitude, and angular velocity, respectively, and generates the motor commands $\mathbf{u}_{r}=[u_{r,1},u_{r,2},u_{r,3},u_{r,4}]^T$. The quadrotor state is then updated from $\mathbf{x}_{r}$ to $\mathbf{x}_{r+1}$ using the actuation and dynamics model in Subsection~\ref{subsec:quadrotor_model}.

The position and velocity errors are
\begin{equation}
\mathbf{e}_{r,\rm q}=\mathbf{q}^{\rm tgt}-\mathbf{q}_{r},\qquad
\mathbf{e}_{r,\rm v}
=
\mathbf{K}_{\rm pos}\mathbf{e}_{r,\rm q}-\mathbf{v}_{r},
\label{eq:pos_vel_error}
\end{equation}
where $\mathbf{K}_{\rm pos}$ is the diagonal position-loop gain matrix.
We decompose the velocity error as $\mathbf{e}_{r,\rm v}=[\mathbf{e}_{r,\rm v,xy}^{T},e_{r,\rm v,z}]^T$ for horizontal and vertical control. For horizontal control, the yaw-decoupled velocity error is mapped to the desired roll $\phi_{r,\rm d}$ and pitch $\theta_{r,\rm d}$ as $[\phi_{r,\rm d},\theta_{r,\rm d}]^T=\mathbf{T}_{\rm map}\mathbf{R}_{\psi_{r}}\mathbf{K}_{\rm vel,xy}\mathbf{e}_{r,\rm v,xy}$, where $\psi_{r}$ is the current yaw angle, $\mathbf{K}_{\rm vel,xy}$ is the diagonal horizontal velocity-loop gain matrix, $\mathbf{R}_{\psi_{r}}$ is the yaw-decoupling matrix, and $\mathbf{T}_{\rm map}$ maps horizontal commands to roll and pitch references. The vertical stream computes the collective throttle as $u_{r,\rm th}=u_{\rm ff}+K_{\rm vel,z}e_{r,\rm v,z}$, where $u_{\rm ff}$ is the gravity-compensation feed-forward term and $K_{\rm vel,z}$ is the vertical velocity-loop gain.

With $\boldsymbol{\Theta}_{r,\rm d}=[\phi_{r,\rm d},\theta_{r,\rm d},\psi^{\rm tgt}]^T$, the desired angular rate and angular-rate correction are
\begin{equation}
\boldsymbol{\omega}_{r,\rm d}
=
\mathbf{K}_{\rm le}(\boldsymbol{\Theta}_{r,\rm d}-\boldsymbol{\Theta}_{r}),\quad
\mathbf{c}_{r,\omega}
=
\mathbf{K}_{\rm lar}(\boldsymbol{\omega}_{r,\rm d}-\boldsymbol{\omega}_{r}),
\label{eq:att_rate_ctrl}
\end{equation}
where $\mathbf{K}_{\rm le}$ and $\mathbf{K}_{\rm lar}$ are the diagonal attitude-loop and angular-rate-loop gain matrices, respectively. Finally, $u_{r,\rm th}$ and $\mathbf{c}_{r,\omega}$ are mixed into the motor command vector as
\begin{equation}
\mathbf{u}_{r}
=
\operatorname{clip}
\left(
u_{r,{\rm th}}\mathbf{1}_{4}
+
\mathbf{M}_{\rm mix}\mathbf{c}_{r,\omega},
0,1
\right),
\label{eq:allocation}
\end{equation}
where $\mathbf{1}_{4}=[1,1,1,1]^T$ and $\mathbf{M}_{\rm mix}$ is the motor-command mixing matrix for the plus configuration.

\begin{figure*}[!t] 
\centering
\captionsetup[subfloat]{font=footnotesize, labelfont=rm, textfont=rm}
\subfloat[]{\includegraphics[width=0.32\linewidth]{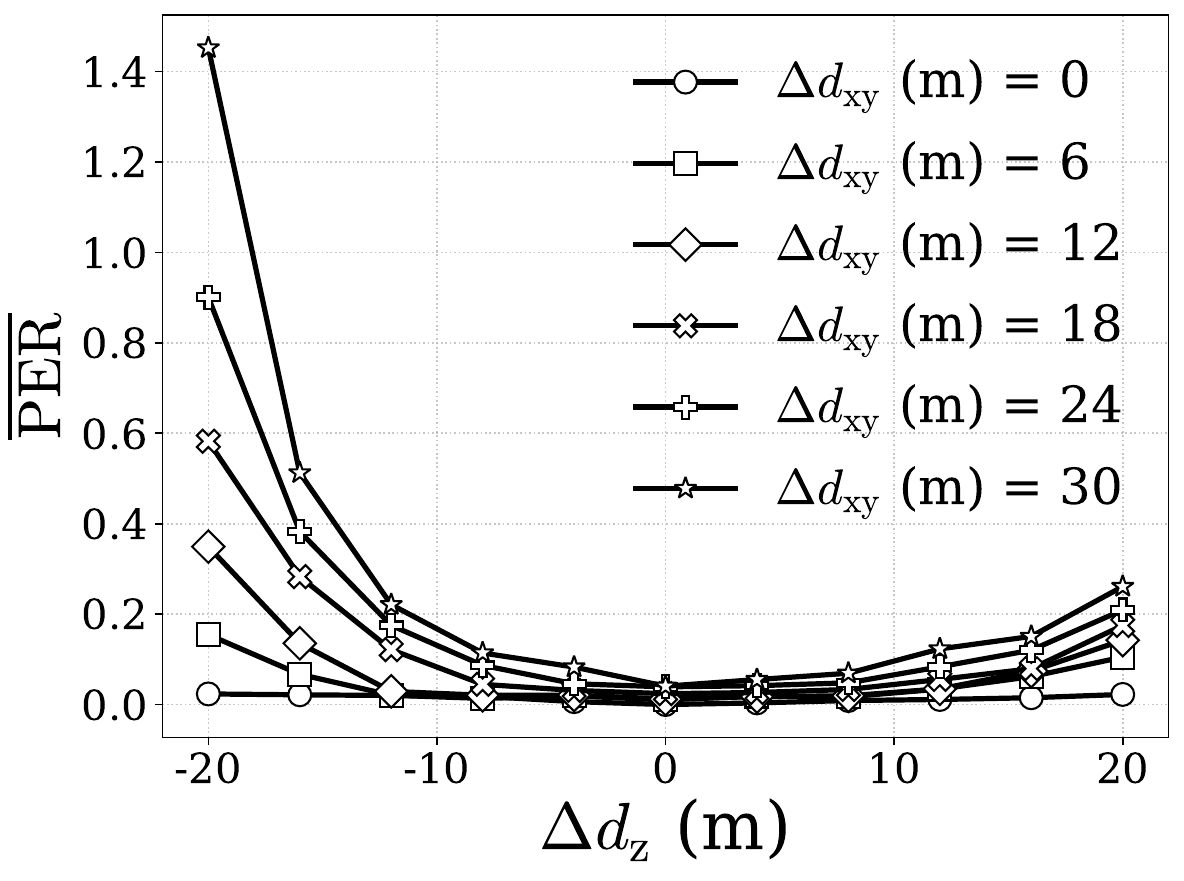}\label{fig:exp1_sec_y}}
\hfil
\subfloat[]{\includegraphics[width=0.32\linewidth]{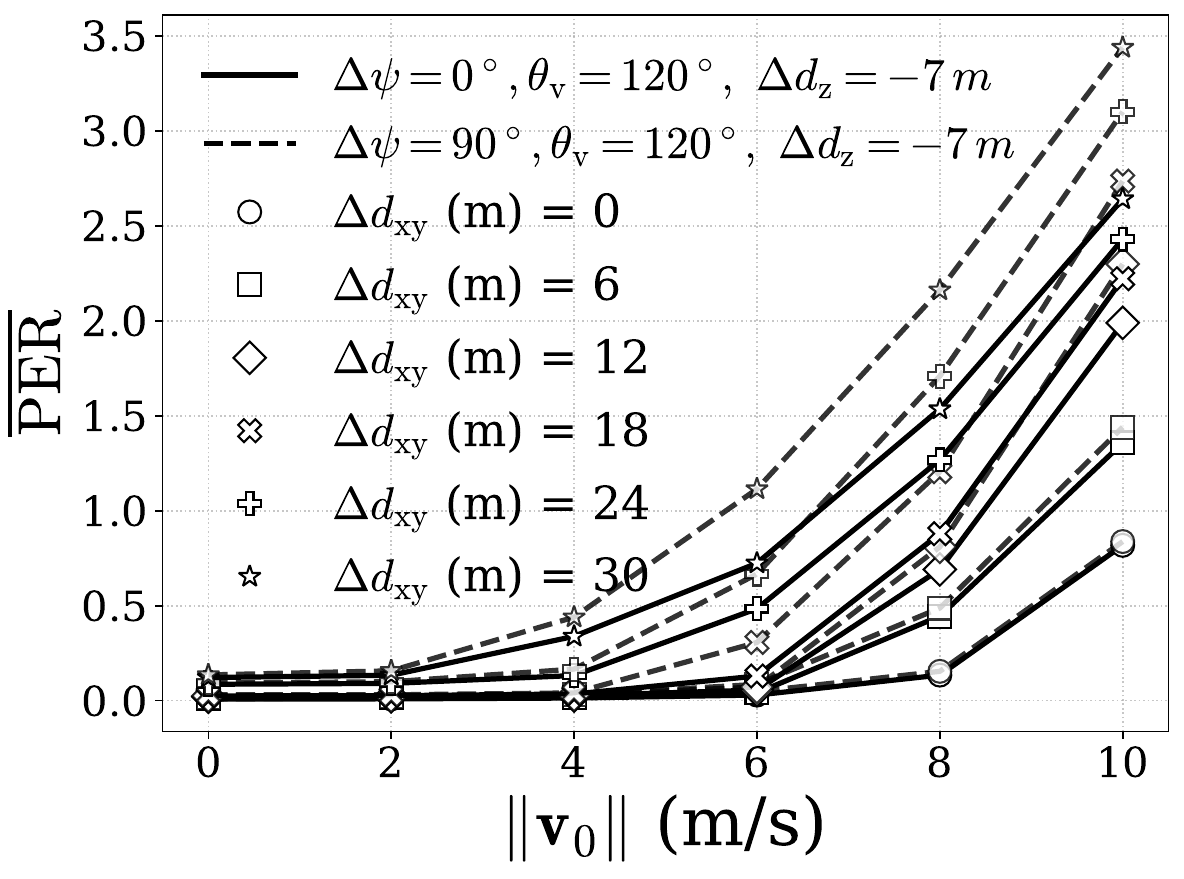}\label{fig:exp4_1}}
\hfil
\subfloat[]{\includegraphics[width=0.32\linewidth]{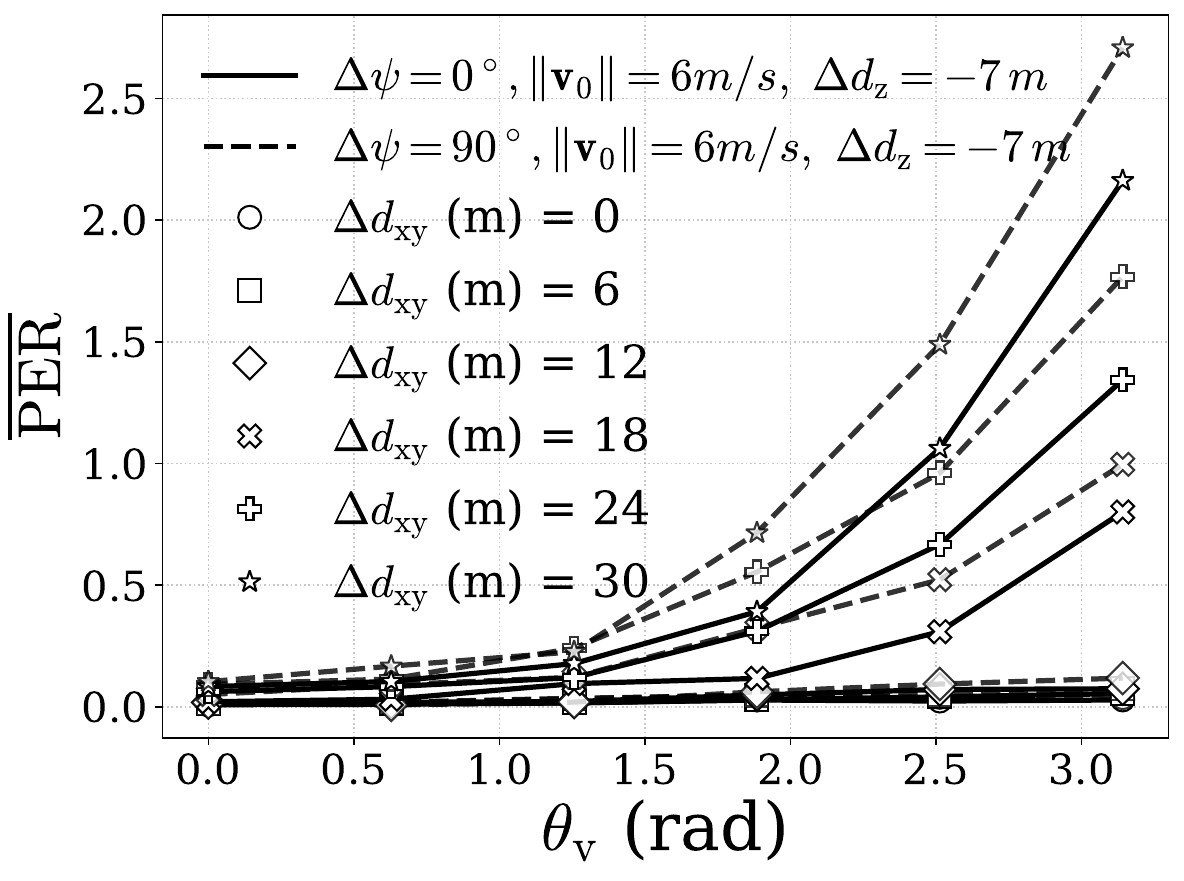}\label{fig:exp4_2}}
\\
\caption{Evaluation of $\overline{\mathrm{PER}}$ under different quadrotor state configurations after finite-horizon execution over $T_{\rm f}$. 
(a) Impact of initial and target positions. 
(b) Impact of initial speed magnitude with and without heading deviation. 
(c) Impact of initial velocity alignment deviation with and without heading deviation.}
\label{fig:double_column_12plots} 
\end{figure*}

\vspace{-0.3cm}
\section{Quadrotor Flight Execution Errors}
\label{Execution_Errors}
As defined in Section~\ref{subsec:control_strategy}, the target position $\mathbf{q}^{\rm tgt}$ is included in the target state $\mathbf{x}^{\rm tgt}$ for finite-horizon flight-control execution. However, $\mathbf{x}^{\rm tgt}$ may not be reached exactly within a finite execution horizon $T_{\rm f}=N_{\rm f}h$. Although such execution errors are constrained by the quadrotor rigid-body dynamics and flight-control algorithms, they are strongly affected by the discrepancy between the initial and target state settings~\cite{romero2022time}.

The 12D quadrotor state evolves in a coupled manner during the $N_{\rm f}$ iterations over $T_{\rm f}$, which may leave residual position, velocity, attitude, and angular-rate errors. The position error directly affects the current communication evaluation by causing the realized position $\mathbf{q}^{\rm act}$ to deviate from $\mathbf{q}^{\rm tgt}$. The quadrotor state after execution over $T_{\rm f}$ becomes the next-slot initial state and further affects subsequent target execution and convergence. Therefore, based on the flight-control model established in Section~\ref{sec:flight_control}, this section characterizes the state-dependent position errors induced by different initial-state and target-state configurations, thereby motivating FM-aware communication optimization.

We first examine how initial and target position settings affect position errors after execution over $T_{\rm f}$. Starting from stable hovering, we vary the horizontal displacement $\Delta d_{\rm xy}=\|\mathbf{q}_{\rm xy}^{\mathrm{tgt}}-\mathbf{q}_{0,\rm xy}\|_2$ and the vertical displacement $\Delta d_{\rm z}=q_{\rm z}^{\mathrm{tgt}}-q_{0,\rm z}$, where the subscripts ${\rm xy}$ and ${\rm z}$ denote horizontal and vertical components, respectively. Positive and negative $\Delta d_{\rm z}$ indicate ascent and descent. The position error ratio (PER) is defined as $\mathrm{PER}=\frac{\|\mathbf{q}^{\mathrm{act}}-\mathbf{q}^{\mathrm{tgt}}\|_2}{\|\mathbf{q}^{\mathrm{tgt}}-\mathbf{q}_{0}\|_2}$, where $\overline{\mathrm{PER}}$ denotes the average PER over multiple experiments.

Fig.~\ref{fig:exp1_sec_y} reveals two observations. First, coupled horizontal-vertical commands yield larger $\overline{\mathrm{PER}}$ than pure horizontal or pure vertical commands, and $\overline{\mathrm{PER}}$ increases with vertical displacement magnitude, since tilt-based horizontal acceleration reduces the thrust margin available for altitude control \cite{hoffmann2007quadrotor}. Second, under the considered flight-control model and actuator limits, descent commands produce larger $\overline{\mathrm{PER}}$ than ascent commands in coupled motion. This is because descent braking requires sufficient upward thrust while the available thrust margin is limited and may approach saturation \cite{ward2022development}.

We then examine how the initial velocity and heading affect $\overline{\mathrm{PER}}$ after execution over $T_{\rm f}$. Let $\|\mathbf{v}_{0}\|$ denote the initial speed magnitude, and let $\theta_{\rm v}$ denote the angle between the initial velocity direction and the desired direction from the initial position to the target position. With $\theta_{\rm v}$ fixed, Fig.~\ref{fig:exp4_1} varies $\|\mathbf{v}_{0}\|$; with $\|\mathbf{v}_{0}\|$ fixed, Fig.~\ref{fig:exp4_2} varies $\theta_{\rm v}$. The heading deviation $\Delta\psi$ is defined as the absolute difference between the initial heading $\psi_{0}$ and the target heading $\psi^{\rm tgt}$, and solid and dashed curves denote the cases without and with heading deviation, respectively. Figs.~\ref{fig:exp4_1} and \ref{fig:exp4_2} show three trends. First, $\overline{\mathrm{PER}}$ increases with $\|\mathbf{v}_{0}\|$, because larger momentum must be dissipated within $T_{\rm f}$ under bounded rotor thrust. Second, $\overline{\mathrm{PER}}$ increases with $\theta_{\rm v}$, since misaligned velocity must be canceled before effective target tracking. Third, heading deviation further increases $\overline{\mathrm{PER}}$, because yaw correction and translational braking share limited actuator authority through attitude control and motor mixing. These trends are consistent with the coupled position and attitude control characteristics of quadrotors under finite thrust and limited attitude control authority~\cite{hoffmann2007quadrotor}.

These observations indicate that finite-horizon quadrotor execution errors are state-dependent, and residual states caused by incomplete convergence in one slot can further affect execution convergence in the next slot. Therefore, the UAV swarm should learn to select communication-efficient deployment positions $\mathbf{q}^{\rm tgt}$ based on the current states, while allowing FM execution to yield realized positions $\mathbf{q}^{\rm act}$ close to $\mathbf{q}^{\rm tgt}$.

\section{Problem Formulation}

\label{problem}
In this section, we formulate a flight-control-constrained joint optimization problem of UAV 3D deployment and power allocation for DL average SE maximization in CF-mMIMO networks. 

At each slot $t$, the SE is evaluated using the realized UAV positions after finite-horizon closed-loop execution. Therefore, SE optimization under the FM model is tightly coupled to the position error $\|\mathbf{q}_{m}^{\rm act}[t]-\mathbf{q}_{m}^{\rm tgt}[t]\|_2$. Reducing this error is necessary to keep the realized UAV positions close to the communication-efficient target positions and preserve the intended channel and power-allocation effects.

Let $\boldsymbol{\rho}=[\rho_{mk}]\in[0,1]^{M\times K}$ denote the DL power allocation ratio matrix, where $\rho_{mk}$ is the fraction of the maximum transmit power $p_m$ allocated by AP $m$ to UE $k$. Accordingly, $p_{mk}^{\rm dl}=\rho_{mk}p_m$. 
The average SE is defined as
$\bar{S}^{\rm dl}(\mathbf{Q}^{\rm act},\boldsymbol{\rho})
=\frac{1}{K}\sum_{k=1}^{K}S_k^{\rm dl}(\mathbf{Q}^{\rm act},\boldsymbol{\rho})$. 
At slot $t$, UAV $m$ is assigned a commanded target position $\mathbf{q}_{m}^{\rm tgt}[t]$. 
After $N_{\rm f}$ closed-loop iterations, the FM execution model in Section~\ref{sec:flight_control} yields the realized position
$\mathbf{q}_{m}^{\rm act}[t]=\Phi_m(\mathbf{x}_{m}[t],\mathbf{q}_{m}^{\rm tgt}[t];T_{\rm f})$. 
For notational simplicity, the index $t$ is omitted hereafter. 

The optimization problem is formulated as
\begin{subequations}\label{P1}
\begin{align}
\mathcal{P}_1:\ 
\max_{\mathbf{Q}^{\rm tgt},\,\mathbf{Q}^{\rm act},\,\boldsymbol{\rho}}\quad
& \bar{S}^{\rm dl}\big(\mathbf{Q}^{\rm act},\boldsymbol{\rho}\big) \label{P1a}\\
\textrm{s.t.}\quad
& \mathbf{q}_{m}^{\rm act}
=\Phi_m\big(\mathbf{x}_{m},\mathbf{q}_{m}^{\rm tgt};T_{\rm f}\big), 
\quad \forall m, \label{P1b}\\
& \mathbf{q}_{m}^{\rm tgt},\,\mathbf{q}_{m}^{\rm act}\in\mathcal{Q},
\quad \forall m, \label{P1c}\\
& \|\mathbf{q}_{m}^{\rm act}-\mathbf{q}_{i}^{\rm act}\|_2\ge d_{\min},
\quad \forall m\neq i, \label{P1d}\\
& \sum_{k=1}^{K}\rho_{mk}\le1,
\quad \forall m, \label{P1e}\\
& 0\le \rho_{mk}\le 1,
\quad \forall m,k. \label{P1f}
\end{align}
\end{subequations}

Constraint~\eqref{P1b} specifies that the realized position $\mathbf{q}_{m}^{\rm act}$ is obtained by the FM model over $T_{\rm f}$. Hence, the optimizer should select target positions whose realized positions remain close to the targets and yield high SE after closed-loop flight-control execution. When \eqref{P1b} is replaced by the ideal relation $\mathbf{q}_{m}^{\rm act}=\mathbf{q}_{m}^{\rm tgt}$, the FM-constrained problem $\mathcal{P}_1$ reduces to the IM-based optimization problem $\mathcal{P}_0$. Constraint~\eqref{P1c} confines both target and realized UAV positions to the feasible deployment region $\mathcal{Q}$. Constraint~\eqref{P1d} imposes a minimum separation $d_{\min}$ among UAV deployment positions. Constraints~\eqref{P1e} and \eqref{P1f} ensure that $\boldsymbol{\rho}$ is a valid per-AP DL power allocation ratio matrix. Problem~$\mathcal{P}_1$ is highly non-convex because it tightly couples SE optimization under complex interference, nonlinear finite-horizon flight-control mapping, and multi-UAV coordination. Therefore, we develop an LLM-enhanced MARL framework to address this problem.

\section{Multi-Agent Reinforcement Learning for Flight-Control-Constrained Joint Optimization}

\label{marl}
\subsection{MARL Formulation}
To address Problem $\mathcal{P}_1$, we model it as a partially observable Markov game, where each UAV acts as an agent. At each slot $t$, agent $m$ selects a continuous action based on its local observation, specifying the deployment and power allocation commands. Under the FM model, target position $\mathbf{q}_m^{\rm tgt}[t]$ is executed through finite flight-control iterations, yielding the realized deployment position $\mathbf{q}_m^{\rm act}[t]$ for DL SE evaluation.

The local observation of agent $m$ is defined as $\mathbf{o}_m[t]=\big[\mathbf{o}_m^{\rm self}[t],\,\mathbf{o}_m^{\rm UE}[t],\,\mathbf{o}_m^{\rm nbr}[t]\big]$. Here, $\mathbf{o}_m^{\rm self}[t]$ contains the UAV self-state features, including boundary distances and the quadrotor flight state. The UE-related observation $\mathbf{o}_m^{\rm UE}[t]$ contains relative geometry, channel gains, and interference values. The neighbor observation $\mathbf{o}_m^{\rm nbr}[t]$ contains relative geometry and service-overlap information with other UAVs. The continuous action of agent $m$ is defined as $\mathbf{a}_m[t]=\big[\mathbf{a}_m^{\rm q}[t],\,\mathbf{a}_m^{\rho}[t]\big]$. $\mathbf{a}_m^{\rm q}[t]\in\mathbb{R}^{3}$ denotes the 3D position command, 
which is mapped to a target position $\mathbf{q}_m^{\rm tgt}[t]\in\mathcal{Q}$. $\mathbf{a}_m^{\rho}[t]\in\mathbb{R}^{K}$ denotes the DL power allocation command. 
The power-allocation action $\mathbf{a}_m^{\rho}[t]$ is mapped into the allocation ratio vector $\boldsymbol{\rho}_m[t]=[\rho_{m1}[t],\ldots,\rho_{mK}[t]]$, 
which satisfies $\sum_{k=1}^{K}\rho_{mk}[t]\le1$. 

\vspace{-0.3cm}
\subsection{MADDPG Training under CTDE}
Following multi-agent deep deterministic policy gradient (MADDPG) \cite{lowe2017multi}, let 
$\boldsymbol{\mu}=\{\mu_1,\ldots,\mu_M\}$ denote the set of deterministic policies of all UAV agents, and let
$\boldsymbol{\xi}=\big[(\mathbf{o}_1)^T,\ldots,(\mathbf{o}_M)^T\big]^T$
denote the global observation vector. Each agent $m$ maintains an actor $\mu_m(\mathbf{o}_m;\theta_m)$, a target actor $\mu'_m(\mathbf{o}_m;\theta'_m)$, a centralized critic 
$Q_m^{\boldsymbol{\mu}}(\boldsymbol{\xi},\mathbf{a}_1,\ldots,\mathbf{a}_M;\phi_m)$, and a target critic 
$Q_m^{\boldsymbol{\mu}'}(\boldsymbol{\xi},\mathbf{a}_1,\ldots,\mathbf{a}_M;\phi'_m)$, where $\theta_m,\phi_m$ and $\theta'_m,\phi'_m$ are the online and target network parameters, respectively. Under the centralized training with decentralized execution (CTDE) paradigm, the critic of agent $m$ uses $\boldsymbol{\xi}$ and the joint action $\mathbf{a}=(\mathbf{a}_1,\ldots,\mathbf{a}_M)$ during training, whereas the actor selects $\mathbf{a}_m=\mu_m(\mathbf{o}_m;\theta_m)$ using only the local observation during decentralized execution.

The replay buffer $\mathcal{D}$ stores joint transitions 
$(\boldsymbol{\xi},\boldsymbol{\xi}',\mathbf{a}_1,\ldots,\mathbf{a}_M,r_1,\ldots,r_M)$, where $r_m$ is the reward of agent $m$ and $\mathbf{r}=(r_1,\ldots,r_M)$. 
The critic of agent $m$ is updated by minimizing
{
\small
\begin{equation}
\mathcal{L}_{\rm c}(\phi_m)=
\mathbb{E}_{\boldsymbol{\xi},\mathbf{a},\mathbf{r},\boldsymbol{\xi}'\sim\mathcal{D}}
\left[
\left(
Q_m^{\boldsymbol{\mu}}(\boldsymbol{\xi},\mathbf{a}_1,\ldots,\mathbf{a}_M;\phi_m)-y_m
\right)^2
\right],
\label{criticloss}
\end{equation}
}
where the target value is
\begin{equation}
y_m
=
r_m+\gamma
Q_m^{\boldsymbol{\mu}'}
(\boldsymbol{\xi}',\mathbf{a}'_1,\ldots,\mathbf{a}'_M;\phi'_m)
\big|_{\mathbf{a}'_j=\mu'_j(\mathbf{o}'_j;\theta'_j)}.
\label{ym}
\end{equation}
Here, $\gamma$ is the discount factor; $\boldsymbol{\xi}'$, $\mathbf{o}'_j$, and $\mathbf{a}'_j$ denote next-slot quantities, while $\mu'_j$, $\theta'_j$, and $\phi'_m$ denote target-network quantities. The actor of agent $m$ is updated using 
\begin{equation}
\begin{aligned}
\nabla_{\theta_m}J(\mu_m)
&=
\mathbb{E}_{\boldsymbol{\xi},\mathbf{a}\sim\mathcal{D}}
\Big[
\nabla_{\theta_m}\mu_m(\mathbf{o}_m;\theta_m) \\
&\hspace{-1.5em}
\nabla_{\mathbf{a}_m}
Q_m^{\boldsymbol{\mu}}(\boldsymbol{\xi},\mathbf{a}_1,\ldots,\mathbf{a}_M;\phi_m)
\big|_{\mathbf{a}_m=\mu_m(\mathbf{o}_m;\theta_m)}
\Big].
\end{aligned}
\label{actorloss}
\end{equation}
The target networks are softly updated as
$\theta'_m\leftarrow\tau_{\rm s}\theta_m+(1-\tau_{\rm s})\theta'_m$ and
$\phi'_m\leftarrow\tau_{\rm s}\phi_m+(1-\tau_{\rm s})\phi'_m$, where $\tau_{\rm s}\ll1$ is the soft-update coefficient.

\vspace{-0.2cm}
\subsection{Motivation for Hybrid Reward Design}
\label{moti}
In the CTDE setting considered here, all UAV agents use a common reward template, but the evaluated reward value differs across agents because their observations and actions are different. Thus, each agent updates its own actor and critic using its corresponding reward signal. Although all UAVs share the global objective of maximizing the average DL SE, a purely global reward provides weak feedback to each agent, whereas purely local rewards may weaken team coordination. Therefore, an effective hybrid reward should balance global components and local components specific to each agent. However, manually designing such rewards and tuning their weights is a tedious process of trial and error and may provide less effective guidance. This motivates LLM-driven reward design, which uses the reasoning and code generation capabilities of LLMs to generate task relevant hybrid rewards for team coordination in MARL.

\begin{figure}[t]
  \centering
  \includegraphics[width=0.98\linewidth]{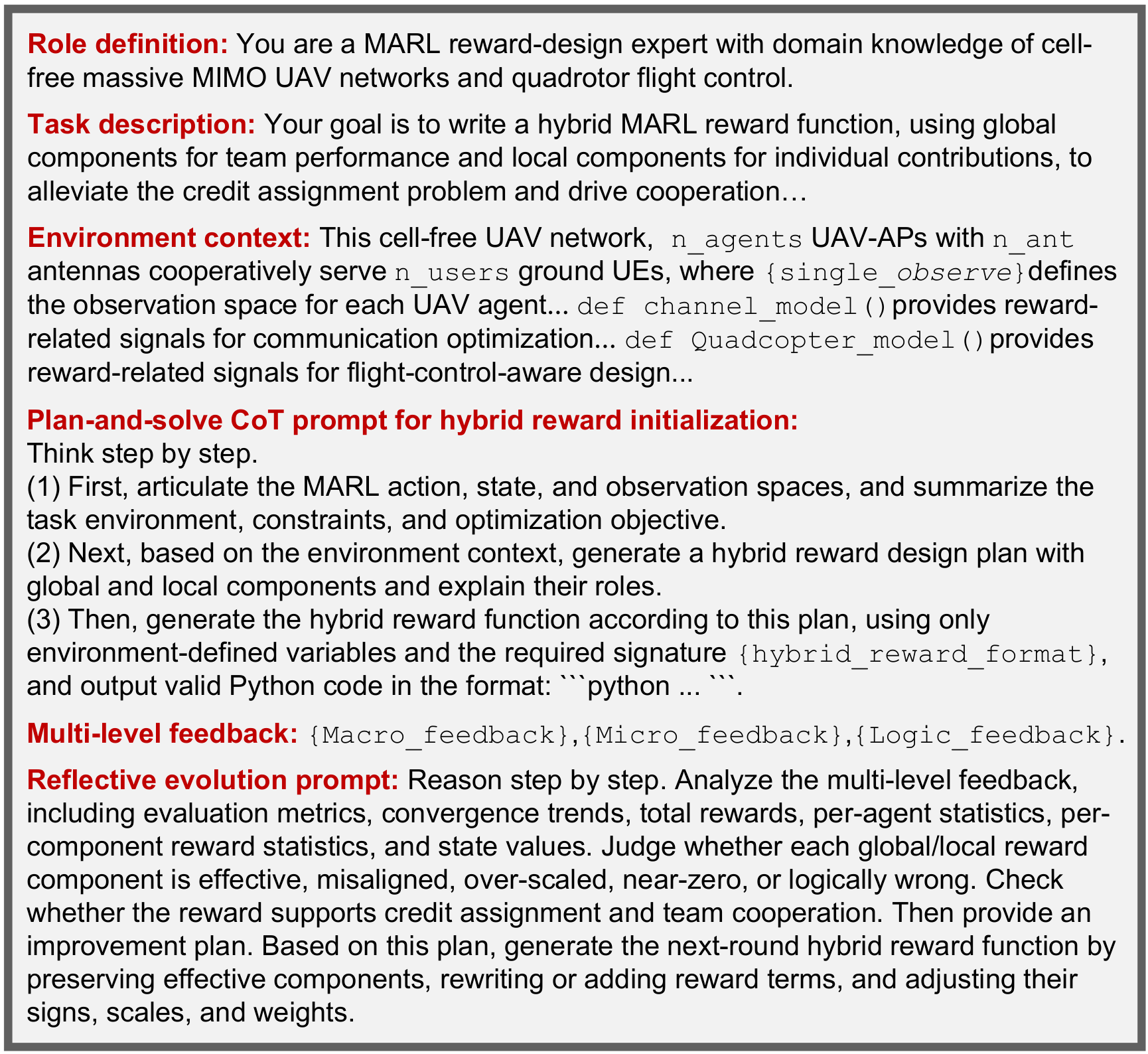}
  \caption{Compact prompt template for hybrid reward evolution.} 
  \label{prompt}
\end{figure}

\begin{algorithm}[t]
\caption{LERE for Solving Problem $\mathcal{P}_1$}
\label{alg:lere_algorithm}
\footnotesize
\begin{algorithmic}[1]
\Require Initial and feedback prompts, MARL environment context, hybrid reward structure, LERE iterations $I=3$, and $C=3$ rewards per iteration.

\State Initialize actor--critic networks, replay buffers, and historical context $\mathcal{H}$.

\For{$i=0,\ldots,I$}
    \If{$i=0$}
        \State Generate $\mathcal{R}_{i}=\{R_{i}^c\}_{c=1}^{C}$ by planner-based initial hybrid rewards.
    \Else
        \State Generate $\mathcal{R}_{i}=\{R_{i}^c\}_{c=1}^{C}$ by reflective evolution with $\mathcal{H}$.
    \EndIf

    \State \parbox[t]{0.82\linewidth} {Apply executability and RLAC-based screening to $\mathcal{R}_{i}$, obtain $\mathcal{R}_{i}^{\rm pass}$, and record logic feedback.}

    \ForAll{$R\in\mathcal{R}_{i}^{\rm pass}$ \textbf{ in parallel}}
        \State Train an independent MADDPG model over episodes using $R$.

        \For{each episode}
            \State Reset the environment and obtain $\{\mathbf{o}_{m}[0]\}_{m=1}^{M}$.

            \For{each slot $t$}
                \State Select $\mathbf{a}_{m}[t]$ and obtain $\mathbf{q}_{m}^{\rm tgt}[t]$ and $\boldsymbol{\rho}_{m}[t]$.
                \State \parbox[t]{0.75\linewidth}{Execute FM over the inter-slot interval $T_{\rm f}$ by Algorithm~\ref{alg:fm_execution} to obtain $\mathbf{q}_{m}^{\rm act}[t]$.}
                \State Compute $\bar{S}^{\rm dl}[t]$ using $\mathbf{Q}^{\rm act}[t]$ and $\boldsymbol{\rho}[t]$.
                \State Store transitions and update networks using Eqs.~\eqref{criticloss}--\eqref{actorloss}.
            \EndFor
        \EndFor

        \State Evaluate the trained policy.
    \EndFor

    \State Select $R_i^{*}$ and construct macro and micro feedback.
    \State \parbox[t]{0.82\linewidth} {Append $R_i^{*}$, logic feedback, macro feedback, micro feedback, and RLAC-failed rewards to $\mathcal{H}$.}
    \State Update $R^{\rm opt}$ according to the evaluation results.
\EndFor

\State \Return $R^{\rm opt}$ and the optimal policy $\boldsymbol{\mu}^{\star}$.
\end{algorithmic}
\end{algorithm}

\begin{figure*}[!t]
  \centering
   \includegraphics[width=18cm]{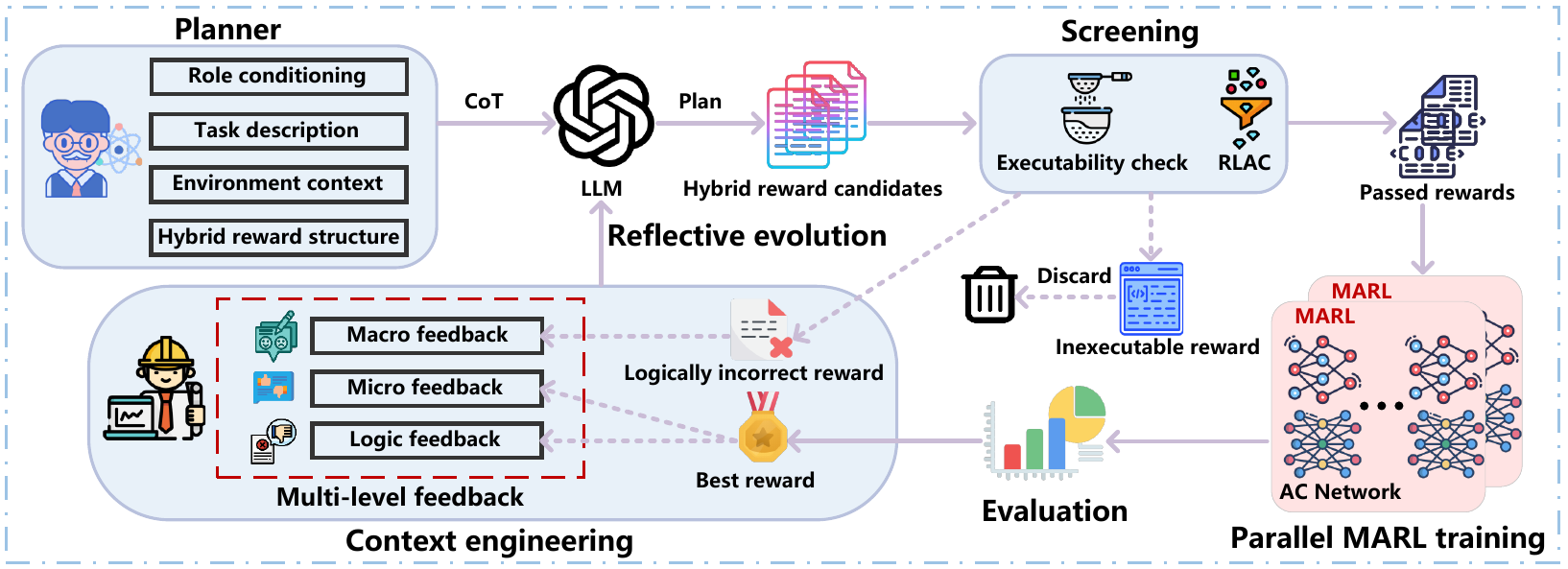}
  \caption{Workflow of the proposed LERE framework.}
  \label{workflow}
\end{figure*}

\section{The Proposed LERE Framework}
\label{sec:lere}
To enable efficient hybrid reward design for MARL, we propose LERE, an LLM-enhanced MARL framework for hybrid reward evolution. For agent $m$ at slot $t$, the hybrid reward template is formulated as 
\begin{equation}
r_m[t]=\sum_i w_{{\rm g},i}G_{i}[t]+\sum_j w_{{\rm l},j}L_{j,m}[t],
\label{hybridreward}
\end{equation}
where $G_{i}[t]$ denotes the $i$-th global reward component shared by all agents, and $L_{j,m}[t]$ denotes the $j$-th local reward component evaluated for agent $m$. The corresponding weights are denoted by $w_{{\rm g},i}$ and $w_{{\rm l},j}$. By evolving both reward components and their weights, LERE balances the global objective and local feedback, thereby promoting multi-UAV cooperation. By invoking the LLM application programming interface (API) in a Python-based environment, LERE automatically synthesizes, screens, evaluates, and evolves hybrid rewards. The prompt template and workflow of LERE are shown in Figs.~\ref{prompt} and~\ref{workflow}, respectively. The complete algorithm for addressing $\mathcal{P}_1$ with LERE is summarized in Algorithm~\ref{alg:lere_algorithm}. The detailed design and workflow of LERE are presented as follows.

1) \textbf{Planner-based hybrid reward initialization:}
The planner first assigns the LLM the role of a MARL reward-design expert and provides the task description. It then presents the MARL environment context using source-code snippets and natural-language explanations, thereby bridging implementation details and task semantics. Given a predefined hybrid reward structure, the planner uses chain-of-thought (CoT) prompting to guide the LLM to derive a hybrid reward plan. Based on this plan, the LLM generates three initial hybrid reward candidates for subsequent screening and evolution.

2) \textbf{Executability and RLAC-based screening:}
Evaluating every generated reward through full MARL training is computationally expensive~\cite{ma2024eureka, li2025efficient}. Therefore, LERE first performs executability screening to discard candidates with syntax errors, invalid weights, or incompatible reward interfaces. Inspired by~\cite{liu2022meta}, we then introduce reward logic alignment check (RLAC) to screen executable rewards before MARL training by verifying whether each candidate hybrid reward assigns higher returns to episodes with better task performance.

We divide the collected offline evaluation episodes into a well-performing set $\mathcal{T}_{\rm well}$ and a poorly performing set $\mathcal{T}_{\rm poor}$ according to task performance. For the multi-agent setting, let $R_i^c$ denote the $c$-th reward function generated in round $i$. The cooperative return of episode $\tau$ under $R_i^c$ is defined as
$\mathcal{G}(\tau|R_i^c)=\frac{1}{M}\sum_{m=1}^{M}\sum_{t=1}^{T} r_m[t]$,
where $r_m[t]$ is the immediate reward of agent $m$ at step $t$, and each episode contains $T$ steps. A candidate reward passes RLAC only if
\begin{equation}
\min_{\tau\in\mathcal{T}_{\rm well}}\mathcal{G}(\tau|R_i^c)
>
\max_{\tau\in\mathcal{T}_{\rm poor}}\mathcal{G}(\tau|R_i^c).
\label{eq:RLAC}
\end{equation}
This conservative condition enforces task-performance consistency by ensuring that episodes with better task performance receive higher cooperative returns. Candidate rewards that fail RLAC are discarded without MARL training. Logic feedback is then generated by comparing the lowest-return episode in $\mathcal{T}_{\rm well}$ with the highest-return episode in $\mathcal{T}_{\rm poor}$, guiding the LLM to revise misaligned reward components and weights in the next iteration.

3) \textbf{Parallel MARL training and evaluation:}
Reward candidates that pass screening are automatically loaded into the MARL reward interface for parallel training. After training, the policy trained with each reward is evaluated under the same criteria. The reward with the best evaluation performance is selected as the best reward in round $i$, denoted by $R_i^{*}$.

4) \textbf{Multi-level feedback:}
Besides logic feedback, LERE constructs macro feedback and micro feedback for $R_i^{*}$. Macro feedback is obtained from periodic episode-level evaluations during training. It summarizes global performance and total returns to reflect team learning, and reports the mean and variance of local metrics and reward components across agents to reveal reward imbalance and contribution disparities.

Since aligning reward values with actual state variables is critical for reward optimization~\cite{li2024auto}, micro feedback is constructed after training from the highest- and lowest-return evaluation episodes. By sampling steps at fixed intervals, it checks whether state variables, such as UAV positions, position errors, and channel gains, are consistent with the global and local reward values. These diagnostics guide the LLM to rebalance global objective and local incentives while identifying and correcting contribution assignment errors.

5) \textbf{Reflective evolution with context engineering:}
Due to the stateless nature of LLMs, they lack persistent memory across independent inference calls and cannot retain a history of previous interactions~\cite{wang2023augmenting}. To support reward evolution over multiple rounds, LERE performs context engineering. After each round, $R_i^{*}$, logic feedback, macro feedback, micro feedback, and RLAC-failed rewards are appended to the historical context $\mathcal{H}$. The historical best reward is explicitly marked as $R^{\rm opt}$. Given this accumulated context, the LLM preserves effective global and local reward structures, revises problematic components, and rebalances weights to generate next round rewards that better align with the global objective.

\section{Experiments and Results}
\label{sec:exp}
This section conducts experiments on the formulated problem using LERE and reward-design baselines. Experimental results show that LERE achieves higher SE, lower position errors, and better reward-design efficiency than the baselines. The results also confirm that IM-trained policies suffer substantial SE degradation under FM execution, highlighting the necessity of FM-aware optimization.

\vspace{-0.3cm}
\subsection{Experimental Setup and Evaluation Protocol}
\label{setup}
The UEs are randomly distributed over the ground service area $\mathcal{A}$. Quadrotor UAVs are initialized with nonzero initial velocities to reflect practical deployment conditions caused by external disturbances, task transitions, or incomplete convergence after previous flight executions. For the LLM-driven reward-design process, all LLM-driven methods call the same GPT-5.4 API with the temperature set to $1$. We adopt episodic training, where each episode contains $T=100$ slots. All compared reward-design methods are trained with the same MADDPG backbone. The actor and critic are fully connected networks with hidden-layer dimensions $\mathbf{l}_{\rm NN}$, and are optimized with learning rates $\eta_{\rm a}$ and $\eta_{\rm c}$, respectively. To stabilize training, minibatches of size $B_{\rm size}$ are sampled from the replay buffer $\mathcal{D}$. Unless otherwise specified, the default simulation parameters are listed in Table~\ref{tab:sim_parameters}.

We compare three reward-design methods, namely, Human, EUREKA~\cite{ma2024eureka, li2025efficient}, and LERE. Human denotes the manually designed reward. EUREKA is a generic LLM-driven evolutionary reward-design framework applicable to MARL. The suffixes -IM and -FM indicate the training environment. Specifically, Human-IM, EUREKA-IM, and LERE-IM are trained under the IM model corresponding to $\mathcal{P}_{0}$, whereas Human-FM, EUREKA-FM, and LERE-FM are trained under the FM model corresponding to $\mathcal{P}_{1}$. Following~\cite{ma2024eureka}, EUREKA generates $16$ candidate rewards per round over one initialization round and five evolution rounds. In contrast, LERE generates $3$ candidate rewards per round over one initialization round and three evolution rounds.

Since different methods use different reward functions, raw reward values are not directly comparable. Therefore, all training and testing results are reported using two metrics at the task level. The first is $\bar{S}_{\rm ep}^{\rm dl}$, defined as the average DL SE over the last five slots of each episode, representing the final optimization performance at the end of the episode. The second metric is $\mathrm{PER}_{\rm ep}$, defined as the average PER over all UAVs and all slots within each episode.

\begin{table}[t!]
\centering
\caption{Default Simulation Parameters}
\label{tab:sim_parameters}
\scriptsize
\setlength{\tabcolsep}{2.4pt}
\renewcommand{\arraystretch}{1.06}
\resizebox{\columnwidth}{!}{
\begin{tabular}{c c c c}
\toprule
\textbf{Symbol} & \textbf{Value} & \textbf{Symbol} & \textbf{Value} \\
\midrule
$M$ & $4$ &
$K$ & $10$ \\

$N$ & $6$ &
$p_m$ & $0.4~{\rm W}$ \\

$\lambda$ & $0.15~{\rm m}$ &
$d_a$ & $\lambda/2$ \\

$f_c$ & $2~{\rm GHz}$ &
$c_0$ & $3\times10^{8}~{\rm m/s}$ \\

$\tau_c$ & $200$ &
$\tau_p$ & $2$ \\

$p_k^{\rm ul}$ & $0.2~{\rm W}$ &
$N_{\rm f}$ & $1200$ \\

$\vartheta$ & $9.61$ &
$\zeta$ & $0.16$ \\

$\eta_{\rm LoS}$ & $1~{\rm dB}$ &
$\eta_{\rm NLoS}$ & $20~{\rm dB}$ \\

$g$ & $9.8~{\rm m/s^2}$ &
$h$ & $0.005~{\rm s}$ \\

$\mathbf{J}$ & ${\rm diag}(1.75,1.75,3.18)\times10^{-2}~{\rm kg\cdot m^2}$ &
$J_{\rm in}$ & $9.90\times10^{-5}~{\rm kg\cdot m^2}$ \\

$C_T$ & $1.25\times10^{-5}$ &
$C_M$ & $1.63\times10^{-7}$ \\

$C_{\varpi}$ & $685~{\rm rad/s}$ &
$\varpi_b$ & $166~{\rm rad/s}$ \\

$\mathbf{K}_{\rm pos}$ & ${\rm diag}(0.42,0.42,0.65)$ &
$\mathbf{K}_{\rm vel,xy}$ & ${\rm diag}(0.22,0.22)$ \\

$K_{\rm vel,z}$ & $0.33$ &
$\mathbf{K}_{\rm le}$ & ${\rm diag}(1.0,1.0,0.91)$ \\

$\mathbf{K}_{\rm lar}$ & ${\rm diag}(0.25,0.25,0.25)$ &
$\mathbf{l}_{\rm NN}$ & $[256,256]$ \\

$\eta_{\rm a}$ & $3\times10^{-4}$ &
$\eta_{\rm c}$ & $1\times10^{-3}$ \\

$\gamma$ & $0.99$ &
$\tau_{\rm s}$ & $0.001$ \\

$\mathcal{D}$ & $50000$ &
$B_{\rm size}$ & $128$ \\

\bottomrule
\end{tabular}}
\end{table}

\begin{table*}[!t]
\centering
\caption{Hybrid Reward Evolution of LERE for Problem $\mathcal{P}_{1}$}
\label{tab:lere_evolution}
\renewcommand{\arraystretch}{1.15}
\setlength{\tabcolsep}{3pt}
\scriptsize
\begin{tabular}{|>{\centering\arraybackslash}p{0.035\textwidth}|
                >{\raggedright\arraybackslash}p{0.09\textwidth}|
                >{\raggedright\arraybackslash}p{0.15\textwidth}|
                >{\raggedright\arraybackslash}p{0.45\textwidth}|
                >{\raggedright\arraybackslash}p{0.17\textwidth}|}
\hline
\textbf{Round} & \textbf{Global terms} & \textbf{Local terms} & \textbf{Feedback summary} & \textbf{Revision focus} \\
\hline

\begin{tabular}[t]{@{}l@{}}
\textbf{Initial} \\
$R_0^{*}$
\end{tabular}
&
\begin{tabular}[t]{@{}l@{}}
\\
$1.3\bar{S}^{\mathrm{dl}}$.
\end{tabular}
&
\begin{tabular}[t]{@{}l@{}}
\\
$-0.5P^{\mathrm{bou}}$, \\
$-0.7P^{\mathrm{F}}$,\\
$-1P^{\mathrm{c}}$.
\end{tabular}
&
\textit{Macro:} SE rises, but inter-agent reward disparity remains high... \newline
\textit{Micro:} UAVs with clearly better position, power-allocation... still receive local rewards close to others, so individual contribution is not well distinguished... \textit{Logic:} \texttt{reward\_init\_2} has a logic error, and the comparison below shows that a failed episode receives higher returns than a successful one...
&
Enhance hybrid global-local incentives and separate agent contributions more clearly. \\
\hline

\begin{tabular}[t]{@{}l@{}}
\textbf{Iter 1} \\
$R_1^{*}$
\end{tabular}
&
\begin{tabular}[t]{@{}l@{}}
\\
$2\bar{S}^{\mathrm{dl}}$, \\
$-0.85P^{\mathrm{interf}}$. 
\end{tabular}
&
\begin{tabular}[t]{@{}l@{}}
\\
$1.2B^{\mathrm{L\mbox{-}SE}}$, $-1.3P^{\mathrm{bou}}$,  \\
$-1P^{\mathrm{F}}$, $-1P^{\mathrm{c}}$. \\
\end{tabular}
&
\textit{Macro:} SE improves, but flight error decreases slowly during training and still fluctuates noticeably... \newline
\textit{Micro:} Across multiple steps with high SE, some UAVs still receive high local rewards despite large velocity, near-boundary motion, or non-negligible position error... \textit{Logic:} \texttt{reward\_iter1\_3} has a logic error...
&
Tighten the coupling between communication gain and action executability. \\
\hline

\begin{tabular}[t]{@{}l@{}}
\textbf{Iter 2} \\
$R_2^{*}$ \\
$R^{\rm opt}$
\end{tabular}
&
\begin{tabular}[t]{@{}l@{}}
\\
$1.8\bar{S}^{\mathrm{dl}}$, \\
$-0.7P^{\mathrm{interf}}$.
\end{tabular}
&
\begin{tabular}[t]{@{}l@{}}
\\
$1.1B^{\mathrm{exec\mbox{-}SE}}$, $-2P^{\mathrm{bar}}$, \\
$-1.5P^{\mathrm{C\mbox{-}F}}$, $-0.8P^{\mathrm{c}}$. \\
\end{tabular}
&
\textit{Macro:} User SE improves, while flight error is reduced to nearly zero... \newline
\textit{Micro:} The reward components of different UAVs are now better aligned with channel quality, power allocation, realized positions, and flight errors, while all agents participate more effectively in cooperation... 
&
Preserve the effective hybrid structure while refining the performance-robustness trade-off. \\
\hline

\begin{tabular}[t]{@{}l@{}}
\textbf{Iter 3} \\
$R_3^{*}$
\end{tabular}
&
\begin{tabular}[t]{@{}l@{}}
\\
$1.3\bar{S}^{\mathrm{dl}}$, \\
$-1.5P^{\mathrm{interf}}$.
\end{tabular}
&
\begin{tabular}[t]{@{}l@{}}
\\
$1.1B^{\mathrm{exec\mbox{-}SE}}$, $-2P^{\mathrm{bar}}$, \\
$-2P^{\mathrm{C\mbox{-}F}}$, $-1P^{\mathrm{c}}$. \\
\end{tabular}
&
\textit{Macro:} SE training becomes more fluctuating, and the test performance falls below the historical best...  
&
Readjust the weights and design reward terms that improve training stability. \\
\hline
\end{tabular}
\end{table*}

\vspace{-0.35cm}
\subsection{Designed Rewards and LERE Evolution Case}
\label{case}
This subsection presents a reward evolution case for Problem $\mathcal{P}_{1}$ to illustrate how LERE evolves hybrid rewards through multi-level feedback. For comparison, we also summarize the Human and EUREKA rewards used under the same MADDPG backbone. Following prior studies on joint UAV deployment and power allocation~\cite{zhong2021multi,xu2023soft}, the Human reward introduces a global SE incentive $\bar{S}^{\rm dl}$ and a collision penalty $P^{\rm col,H}$. For a fair FM-aware comparison specific to $\mathcal{P}_{1}$, Human reward further adds a position error penalty $\|\mathbf{q}^{\mathrm{act}}-\mathbf{q}^{\mathrm{tgt}}\|_2$. The EUREKA reward is evolved using the generic LLM-driven reward-design method~\cite{ma2024eureka}. Compared with the Human reward, it introduces a global channel-gain incentive 
$\bar{\beta}=\frac{1}{K}\sum_{k=1}^{K}\frac{1}{M}\sum_{m=1}^{M}\beta_{mk}$ 
and a global effective-interference penalty 
$\bar{I}^{\rm dl}=\frac{1}{K}\sum_{k=1}^{K}I_k^{\rm dl}$, 
where $I_k^{\rm dl}$ denotes the non-noise interference term in the denominator of Eq.~\eqref{eq:dl_sinr}. Hard boundary penalties are also added to guide UAV deployment within the feasible region during the early stages of training.

Table~\ref{tab:lere_evolution} summarizes how LERE evolves the global and local terms of the hybrid reward for Problem~$\mathcal{P}_{1}$. The initial hybrid reward contains the global SE term $\bar{S}^{\rm dl}$ and simple local terms, including boundary penalty $P^{\rm bou}$, PER penalty $P^{\rm F}$, and collision penalty $P^{\rm c}$. For UAV $m$, $P^{\rm F}$ denotes the PER penalty for finite-horizon position error. Although this initial reward improves SE, micro feedback reveals weak alignment between rewards and states, where individual UAV contributions are not well reflected by their local returns. RLAC also detects a reward logic error, and LERE generates logic feedback accordingly.

Based on these feedback signals, Iter~1 adds the global interference penalty $P^{\rm interf}$ and the local SE contribution term $B^{\rm L\mbox{-}SE}$. The term $P^{\rm interf}$ suppresses inter-user interference, where $P^{\rm interf}=\frac{1}{K}\sum_{k=1}^{K}\log_{10}(1+I_k)$ and $I_k=\sum_{l\neq k}\mathbb{E}\{|\sum_{i=1}^{M}\mathbf{f}_{il}^{H}\mathbf{g}_{ik}|^{2}\}$. The term $B^{\rm L\mbox{-}SE}=\frac{\sum_{k=1}^{K}\chi_{mk}S_k^{\rm dl}}{\sum_{k=1}^{K}S_k^{\rm dl}+\epsilon}$ assigns SE contribution to UAV $m$ according to its channel strength and allocated DL power, where $\chi_{mk}=\frac{p_{mk}^{\rm dl}\beta_{mk}}{\sum_{i=1}^{M}p_{ik}^{\rm dl}\beta_{ik}+\epsilon}$. However, micro feedback shows that some UAVs still receive high local rewards despite non-negligible position errors, indicating that the local SE contribution is not sufficiently linked to FM-induced position errors.

Therefore, Iter~2 replaces $B^{\rm L\mbox{-}SE}$ with the flight-error-coupled term 
$B^{\rm exec\mbox{-}SE}=B^{\rm L\mbox{-}SE}\exp(-\eta_{\rm F}P^{\rm C\mbox{-}F})$, 
replaces the hard boundary penalty with the barrier penalty $P^{\rm bar}$, and introduces the state-related PER penalty $P^{\rm C\mbox{-}F}$. 
A compact form is 
$P^{\rm C\mbox{-}F}=P^{\rm F}\left(1+\eta_{\rm v}\frac{\|\mathbf{v}\|_2}{v_{\max}}+\eta_{\theta}(1-\cos\theta_{\rm v})\right)$, 
where $\eta_{\rm F}$, $\eta_{\rm v}$, and $\eta_{\theta}$ are positive shaping coefficients, and $v_{\max}$ is used for speed normalization. 
This term imposes stronger penalties on UAVs with higher speed or poorer velocity-direction alignment. 
Feedback indicates improved team cooperation and effective individual incentives, suggesting that subsequent evolution can focus on reward weight refinement. In Iter~3, macro feedback shows that the evaluation performance of the policy trained with $R_3^{*}$ falls below that trained with $R_2^{*}$. Therefore, $R_2^{*}$ is retained as the final historical best reward $R^{\rm opt}$.

Overall, this reward evolution case shows that LERE structurally evolves hybrid reward functions that support team cooperation and effective individual incentives. Its RLAC, macro-feedback, and micro-feedback modules guide the LLM to check reward logic, assess training behavior, and align reward values with states. Therefore, LERE shows stronger potential than the baselines for multi-agent cooperative tasks.

\begin{figure}[!t]
  \centering
  \includegraphics[width=0.8\linewidth]{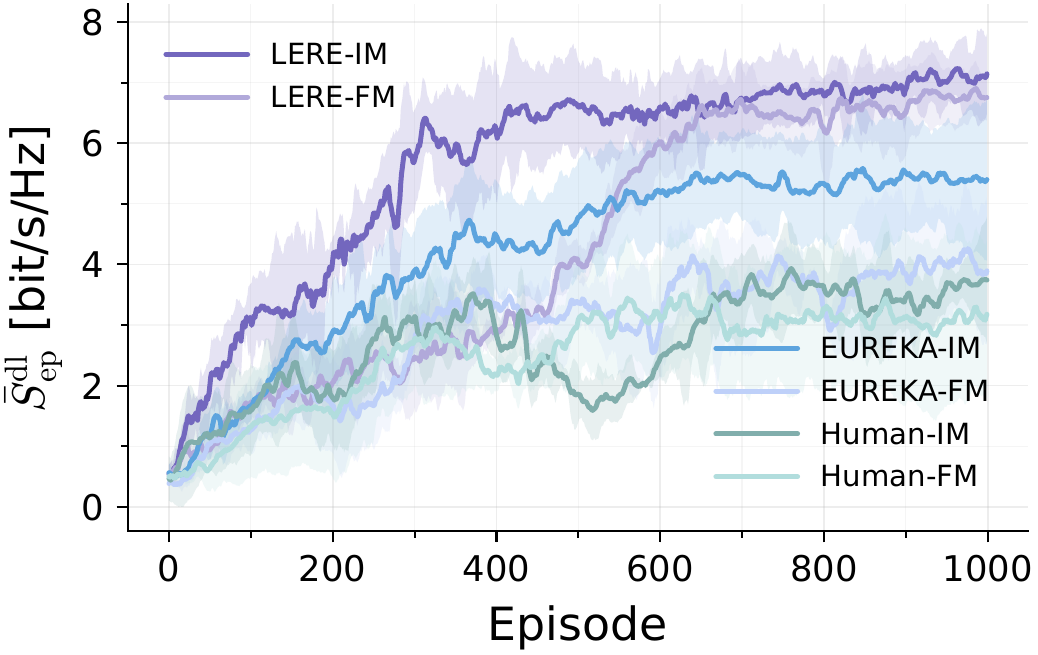}
  \caption{Convergence of $\bar{S}_{\rm ep}^{\rm dl}$ under IM and FM training models.}
  \label{train-se}
\end{figure}

\begin{figure}[!t]
  \centering
  \includegraphics[width=0.8\linewidth]{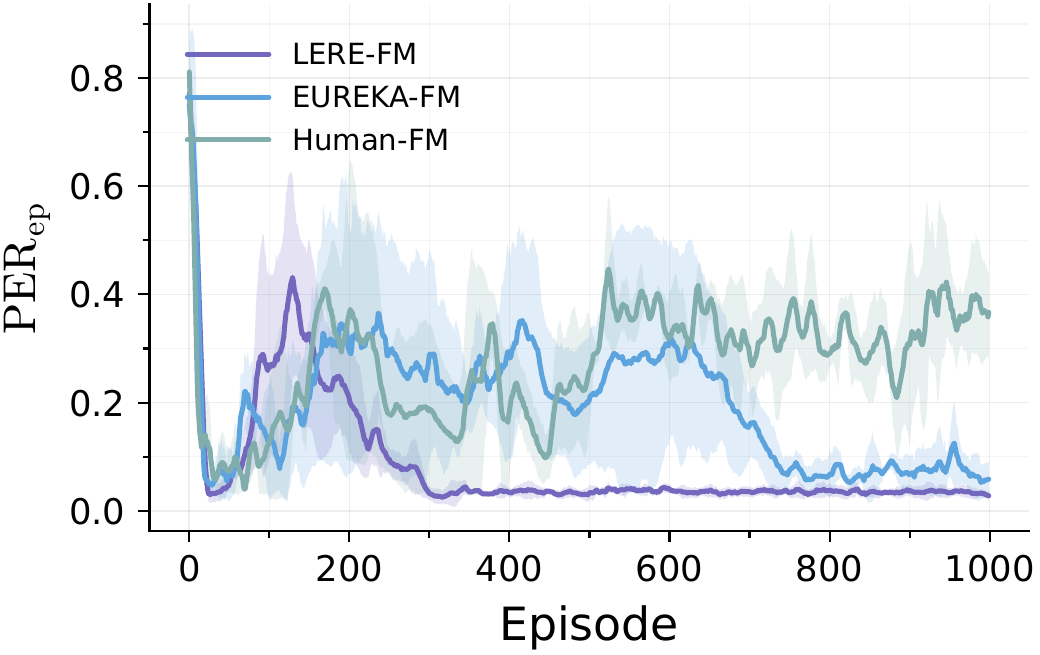}
  \caption{Convergence of $\mathrm{PER}_{\rm ep}$ under the FM training model.}
  \label{train-per} 
\end{figure}

\vspace{-0.2cm}
\subsection{Training Results under IM and FM Models}
\label{train}

As shown in Fig.~\ref{train-se}, LERE achieves the highest training $\bar{S}_{\rm ep}^{\rm dl}$ among the compared reward-design methods under both IM and FM training models. Under the IM training model, UAVs are assumed to exactly reach the commanded positions, so $\mathrm{PER}_{\rm ep}$ is zero for all methods and is not plotted in Fig.~\ref{train-per}. LERE-IM converges faster and achieves a higher final $\bar{S}_{\rm ep}^{\rm dl}$ than EUREKA-IM and Human-IM, indicating that the LERE-evolved reward provides effective guidance for ideal-motion SE optimization.

Under the FM training model, finite-horizon quadrotor execution introduces position errors, making policy learning more difficult. Although the FM curves fluctuate during early exploration, LERE-FM achieves a higher final $\bar{S}_{\rm ep}^{\rm dl}$ than EUREKA-FM and Human-FM. Moreover, Fig.~\ref{train-per} shows that LERE-FM gradually suppresses $\mathrm{PER}_{\rm ep}$ and stabilizes at a much lower level, whereas EUREKA-FM and Human-FM retain non-negligible position errors after training. These results show that the LERE-evolved reward balances global SE optimization and individual UAV flight constraints, enabling each UAV to learn communication-efficient and accurately executable deployment commands.

\subsection{Matched and Mismatched Execution Tests}
\label{test}
We evaluate the trained policies under matched and mismatched UAV motion execution models. The results show that LERE achieves the best matched-test performance, while IM-trained policies suffer clear SE degradation when executed under the FM model.

\begin{figure}[b]
  \centering
  \includegraphics[width=0.8\linewidth]{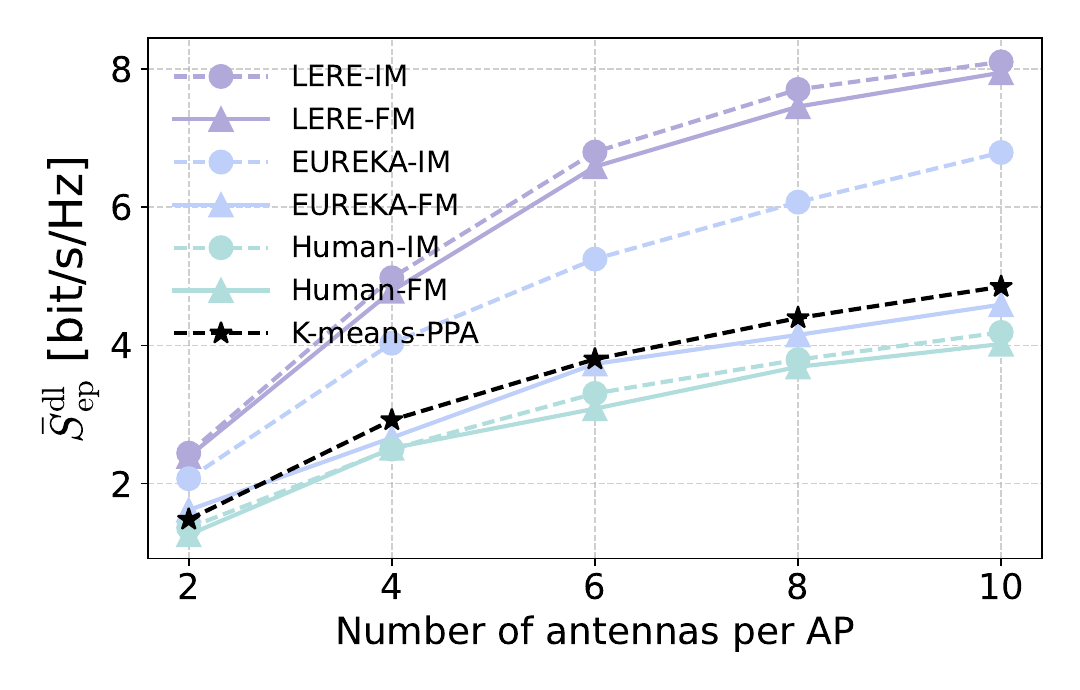}
  \caption{Matched-test $\bar{S}_{\rm ep}^{\rm dl}$ versus antenna number.}
  \label{test-1} 
\end{figure}

In the matched tests, each policy is evaluated under the same motion model used for training. Fig.~\ref{test-1} shows the matched-test $\bar S_{\rm ep}^{\rm dl}$ versus the antenna number $N$. K-means-PPA is also included as a non-MARL baseline, where UAV deployment is obtained by K-means clustering \cite{kanungo2002efficient} and DL power is allocated in proportion to the channel quality \cite{ozdogan2019performance}. The results show that LERE-IM and LERE-FM achieve the highest $\bar{S}_{\rm ep}^{\rm dl}$ under the IM and FM execution tests, respectively, outperforming the other compared methods. This shows that the LERE-evolved reward achieves strong matched-test performance under both IM and FM execution and consistently performs well over the tested antenna numbers.

\begin{figure}[b]
  \centering
  \includegraphics[width=0.8\linewidth]{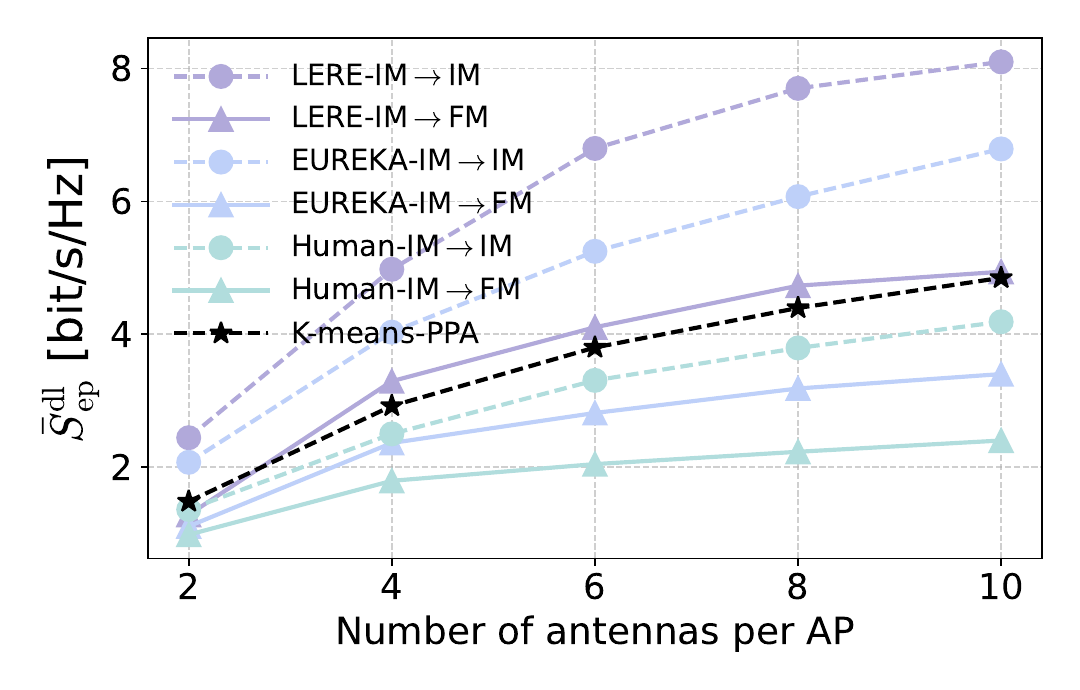}
  \caption{Mismatched-test $\bar{S}_{\rm ep}^{\rm dl}$ versus antenna number.}
  \label{test-2}
\end{figure}

Fig.~\ref{test-2} further evaluates IM-trained policies under mismatched FM execution tests versus the antenna number $N$. For all reward-design methods, $\bar S_{\rm ep}^{\rm dl}$ clearly decreases from IM$\rightarrow$IM to IM$\rightarrow$FM, indicating that policies optimized under ideal point-mass motion cannot be directly executed under FM without performance loss.

Specifically, Table~\ref{tab:cross_eval} reports the test results at $N=6$, where $\bar{S}_{\rm ep,IM}^{\rm dl}$ and $\bar{S}_{\rm ep,FM}^{\rm dl}$ denote $\bar{S}_{\rm ep}^{\rm dl}$ evaluated under IM and FM execution tests, respectively. $\Delta_{\rm drop}$ measures the relative $\bar{S}_{\rm ep}^{\rm dl}$ loss when an IM-trained policy is tested under the FM model, while $\Delta_{\rm gain}$ measures the relative $\bar{S}_{\rm ep}^{\rm dl}$ improvement of an FM-trained policy over its IM-trained counterpart under FM execution tests. For IM-trained policies, switching from IM to FM execution causes substantial $\bar{S}_{\rm ep}^{\rm dl}$ degradation, with $\Delta_{\rm drop}$ values of $37.88\%$, $46.37\%$, and $39.62\%$ for Human-IM, EUREKA-IM, and LERE-IM, respectively, accompanied by large $\mathrm{PER}_{\rm ep}$ values. In contrast, FM-trained policies achieve higher $\bar{S}_{\rm ep}^{\rm dl}$ than their IM-trained counterparts under FM execution tests. In particular, LERE-FM increases $\bar{S}_{\rm ep,FM}^{\rm dl}$ from $4.10$ to $6.58$ bit/s/Hz compared with LERE-IM, corresponding to a $\Delta_{\rm gain}$ of $60.49\%$, while reducing $\mathrm{PER}_{\rm ep}$ from $0.39$ to $0.04$. These results show that IM-trained policies suffer substantial performance loss under FM execution tests, highlighting the necessity of flight-control-constrained modeling and FM-aware training for maintaining communication performance under FM execution.

\begin{figure}[t!]
  \centering
  \includegraphics[width=0.76\linewidth]{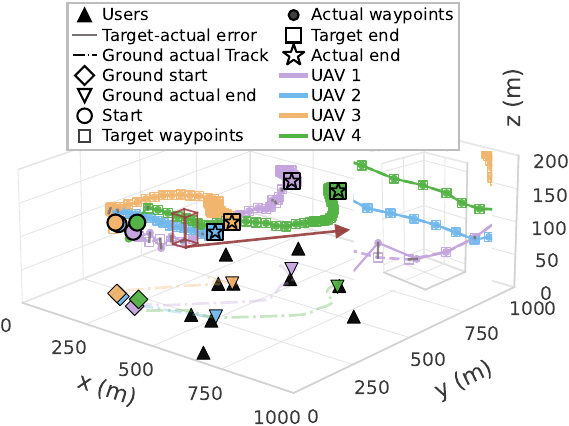}
  \caption{Trajectory test of LERE-FM under FM execution.}
  \label{test-traj1} 
\end{figure}

\begin{figure}[t!]
  \centering
  \includegraphics[width=0.76\linewidth]{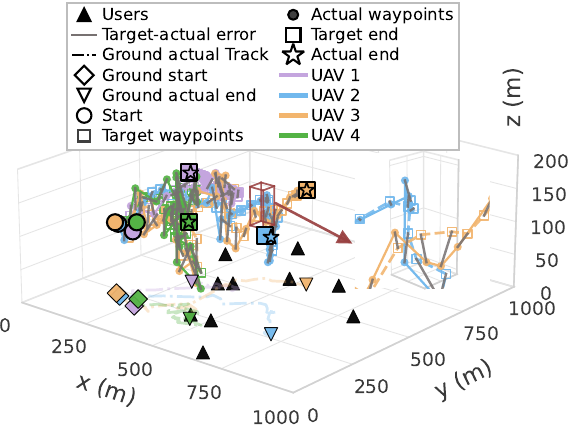}
  \caption{Trajectory test of LERE-IM under FM execution.}
  \label{test-traj2} 
\end{figure}

To intuitively explain the performance degradation observed in the mismatched tests, Fig.~\ref{test-traj1} and Fig.~\ref{test-traj2} show the FM-executed test trajectories of the LERE policies trained under FM and IM, respectively, using the same test environment configuration. LERE-FM produces realized trajectories that closely follow the target trajectories and achieves $\bar{S}_{\rm ep}^{\rm dl}=6.73$ bit/s/Hz. In contrast, LERE-IM exhibits larger deviations between target and realized trajectories and only achieves $\bar{S}_{\rm ep}^{\rm dl}=4.06$ bit/s/Hz. These deviations make the realized channels between UAVs and UEs differ from those expected at the target positions, thereby weakening the effectiveness of the power allocation designed for the target deployment and causing SE degradation.

\begin{table}[t]
\centering
\caption{Matched and Mismatched Test Results at $N=6$}
\label{tab:cross_eval}
\small
\setlength{\tabcolsep}{4pt}
\begin{tabular}{lccccc}
\toprule
\textbf{Policy} & \textbf{$\bar{S}_{\rm ep,IM}^{\rm dl}$} & \textbf{$\bar{S}_{\rm ep,FM}^{\rm dl}$} & \textbf{$\mathrm{PER}_{\rm ep}$} & \textbf{$\Delta_{\mathrm{drop}}$} & \textbf{$\Delta_{\mathrm{gain}}$} \\
\midrule
\textbf{Human-IM}      & 3.30  & 2.05  & 0.61  & $37.88\%$  & --             \\
\textbf{Human-FM}   & --            & 3.08  & 0.36  & --            & $50.24\%$  \\
\textbf{EUREKA-IM}     & 5.24  & 2.81  & 0.52  & $46.37\%$  & --             \\
\textbf{EUREKA-FM}  & --            & 3.73  & 0.14  & --            & $32.74\%$  \\
\textbf{LERE-IM}       & 6.79  & 4.10  & 0.39  & $39.62\%$  & --             \\
\textbf{LERE-FM}    & --            & 6.58  & 0.04  & --            & $60.49\%$  \\
\bottomrule
\end{tabular}
\end{table}

\subsection{Ablation Study and Reward-Design Efficiency}
\label{Ablation}
We conduct ablation studies under FM training and FM execution testing to identify LERE's performance gains for Problem~$\mathcal{P}_1$. The results show that the gain comes from the overall design of the LERE framework rather than a single module. We further analyze reward-design efficiency and show that LERE achieves higher efficiency than EUREKA.

\subsubsection{Ablation Study}
Table~\ref{tab:ablation} reports the ablation results for the reward structure and feedback modules. Full LERE uses the complete optimal reward $R^{\rm opt}$ evolved at Iter~2 in Table~\ref{tab:lere_evolution}. Only Global and Only Local retain only the global terms and local terms of $R^{\rm opt}$, respectively. For the ablations of the feedback modules, w/o Macro, w/o Micro, and w/o Logic denote the hybrid rewards evolved by LERE after removing macro feedback, micro feedback, and logic feedback, respectively.

The results show that Full LERE achieves a higher $\bar{S}_{\rm ep}^{\rm dl}$ than all ablated variants. Only Global obtains a low $\bar{S}_{\rm ep}^{\rm dl}$ and a large $\mathrm{PER}_{\rm ep}$ because the shared team reward lacks guidance specific to each UAV. Only Local maintains a low $\mathrm{PER}_{\rm ep}$ but achieves a lower $\bar{S}_{\rm ep}^{\rm dl}$, showing that local incentives alone cannot fully optimize the global objective. Among the feedback modules, removing micro feedback causes the largest degradation, indicating that step-level feedback for aligning state and reward values is necessary. The performance degradation caused by removing macro or logic feedback further indicates that the gain of LERE comes from the overall framework rather than from a single module.

Table~\ref{tab:planner_context_ablation} evaluates the effect of context engineering during reward evolution. Compared with LERE, w/o Context removes the historical memory context and provides only the current-iteration feedback to the LLM. It reaches a lower peak $\bar{S}_{\rm ep}^{\rm dl}$ of $5.84$ compared with $6.58$ for LERE and drops sharply at Iter~3, showing that context engineering improves the evolved reward performance and helps mitigate severe performance regression.

\subsubsection{Efficiency Comparison}
We further compare the reward-design efficiency of LERE and EUREKA over their complete reward-design processes. LERE uses only $12$ LLM queries, whereas EUREKA uses $96$ LLM queries following~\cite{ma2024eureka}. LERE consumes $184{,}617$ tokens, whereas EUREKA consumes $2{,}269{,}374$ tokens. The executability rate is defined as the percentage of generated reward candidates that can be successfully parsed and embedded into the MARL environment. LERE achieves a higher executability rate than EUREKA, namely $90.2\%$ versus $71.4\%$. Moreover, as shown in Table~\ref{tab:cross_eval}, LERE-FM achieves $\bar{S}_{\rm ep}^{\rm dl}=6.58$ and $\mathrm{PER}_{\rm ep}=0.04$, while EUREKA-FM achieves only $\bar{S}_{\rm ep}^{\rm dl}=3.73$ and $\mathrm{PER}_{\rm ep}=0.14$. This indicates that LERE produces a more effective reward at a lower reward-design cost than EUREKA.

\begin{table}[t!]
\centering
\caption{Reward Structure and Feedback Ablation}
\label{tab:ablation}
\setlength{\tabcolsep}{5pt} 
\begin{tabular}{lcc lcc}
\toprule
\textbf{Variant} & \textbf{$\bar{S}_{\rm ep}^{\rm dl}$} & \textbf{$\mathrm{PER}_{\rm ep}$} &
\textbf{Variant} & \textbf{$\bar{S}_{\rm ep}^{\rm dl}$} & \textbf{$\mathrm{PER}_{\rm ep}$} \\
\midrule
\textbf{Full LERE}   & 6.58 & 0.04 & \textbf{w/o Macro} & 5.93 & 0.06 \\
\textbf{Only Global} & 1.79 & 0.76 & \textbf{w/o Micro} & 3.92 & 0.18 \\
\textbf{Only Local}  & 4.35 & 0.03 & \textbf{w/o Logic} & 5.36 & 0.07 \\
\bottomrule
\end{tabular}
\end{table}

\begin{table}[t!]
\caption{Reward-Evolution Ablation}
\label{tab:planner_context_ablation}
\centering
\begin{tabular}{lccccc}
\toprule
\textbf{Variant} & \textbf{Metric} & \textbf{Initial} & \textbf{Iter 1} & \textbf{Iter 2} & \textbf{Iter 3} \\
\midrule
\multirow{2}{*}{\textbf{LERE}} & \textbf{$\bar{S}_{\rm ep}^{\rm dl}$} & 2.74  & 4.32 & 6.58 & 6.29  \\
 & \textbf{$\mathrm{PER}_{\rm ep}$} & 0.26 & 0.07 & 0.04 & 0.04 \\
\midrule
\multirow{2}{*}{\textbf{w/o Context}} & \textbf{$\bar{S}_{\rm ep}^{\rm dl}$} & 2.79 & 3.98 & 5.84 & 4.26 \\
 & \textbf{$\mathrm{PER}_{\rm ep}$} & 0.23 & 0.14 & 0.06 & 0.08 \\
\bottomrule
\end{tabular}
\end{table}

\vspace{-0.3cm}
\subsection{Reward Transferability Across Task Scales}
\begin{figure}[!t]
  \centering
  \includegraphics[width=0.8\linewidth]{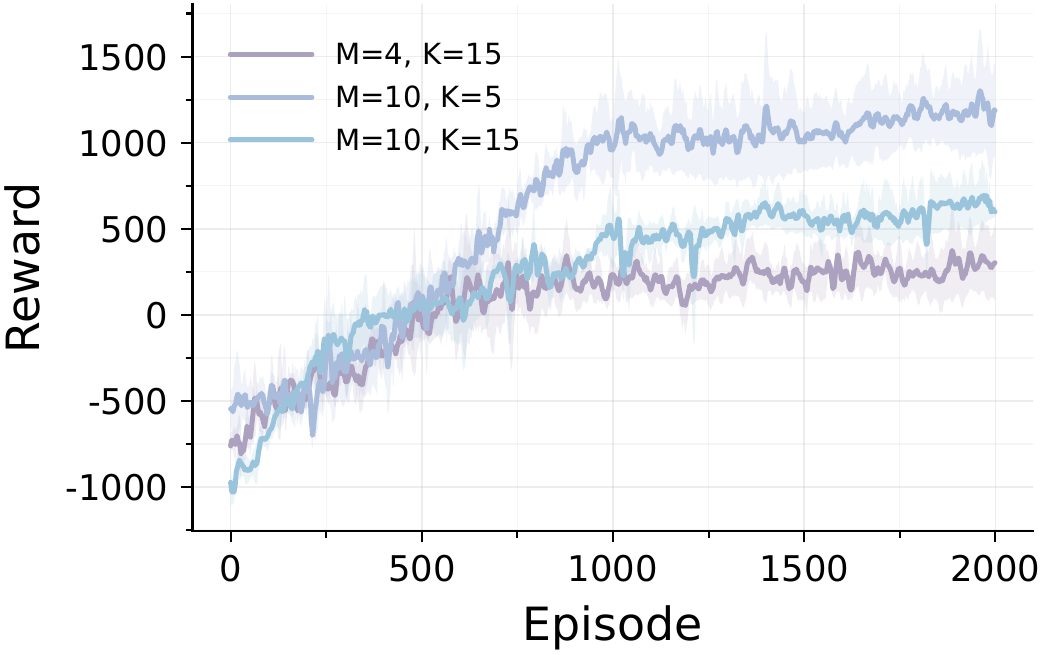}
  \caption{LERE reward convergence under different task scales.}
  \label{fig:reward_transfer}
\end{figure}

Finally, we examine whether the reward evolved by LERE remains effective when the task scale changes. We reuse the same optimal reward $R^{\rm opt}$ evolved for problem $\mathcal{P}_{1}$, as shown in Table~\ref{tab:lere_evolution}, and retrain MARL policies with $N=6$ under different $(M,K)$ settings. Fig.~\ref{fig:reward_transfer} reports the mean episodic return. Since reward magnitudes vary with the number of UAVs and UEs, we focus on convergence trends rather than absolute values. All curves increase and gradually stabilize, suggesting that $R^{\rm opt}$ can still provide useful learning guidance across different task scales.

\label{Scalability}

\section{Conclusion}
This paper formulated a flight-control-constrained joint optimization problem for 3D UAV deployment and DL power allocation in CF-mMIMO networks. By modeling each UAV as a 6-DoF quadrotor with finite-horizon closed-loop flight-control execution, the proposed formulation evaluates SE at realized UAV positions, thereby explicitly accounting for position errors. To address this problem, we proposed LERE, an LLM-enhanced MARL framework that evolves hybrid rewards for team coordination. Experimental results showed that LERE achieved the highest SE across different motion execution models. Under FM execution tests, FM-trained policies significantly outperformed their IM-trained counterparts, while IM-trained policies suffered clear SE degradation due to deviations between target and realized positions. Ablation results verified that the performance gain of LERE comes from the full framework, and the efficiency comparison showed that LERE produced more effective rewards at a lower design cost. These findings highlight the necessity of considering flight-control constraints in UAV communication optimization, particularly for SE optimization, and show that LERE effectively enhances MARL for multi-UAV cooperative tasks.

\bibliographystyle{IEEEtran}
\bibliography{IEEEabrv,refer}

\begin{thebibliography}{10}
\providecommand{\url}[1]{#1}
\csname url@samestyle\endcsname
\providecommand{\newblock}{\relax}
\providecommand{\bibinfo}[2]{#2}
\providecommand{\BIBentrySTDinterwordspacing}{\spaceskip=0pt\relax}
\providecommand{\BIBentryALTinterwordstretchfactor}{4}
\providecommand{\BIBentryALTinterwordspacing}{\spaceskip=\fontdimen2\font plus
\BIBentryALTinterwordstretchfactor\fontdimen3\font minus \fontdimen4\font\relax}
\providecommand{\BIBforeignlanguage}[2]{{%
\expandafter\ifx\csname l@#1\endcsname\relax
\typeout{** WARNING: IEEEtran.bst: No hyphenation pattern has been}%
\typeout{** loaded for the language `#1'. Using the pattern for}%
\typeout{** the default language instead.}%
\else
\language=\csname l@#1\endcsname
\fi
#2}}
\providecommand{\BIBdecl}{\relax}
\BIBdecl

\bibitem{elhoushy2021cell}
S.~Elhoushy, M.~Ibrahim, and W.~Hamouda, ``Cell-free massive mimo: A survey,'' \emph{IEEE Commun. Surv. Tutorials}, vol.~24, no.~1, pp. 492--523, 2022.

\bibitem{shah2026joint}
S.~A.~A. Shah, X.~N. Fernando, and R.~Kashef, ``Joint optimization of uav trajectory, transmit power, and user association in aerial-terrestrial cell-free massive mimo network,'' \emph{IEEE Trans. Wireless Commun.}, vol.~25, pp. 15\,818--15\,832, 2026.

\bibitem{wan2024performance}
Z.~Wan, J.~Li, P.~Zhu, D.~Wang, F.~Liu, and X.~You, ``Performance analysis of multi-uav aided cell-free radio access network with network-assisted full-duplex for urllc,'' \emph{IEEE Trans. Commun.}, vol.~72, no.~9, pp. 5810--5822, Sep. 2024.

\bibitem{shi2022meta}
M.~Shi, K.~Yang, D.~Niyato, H.~Yuan, H.~Zhou, and Z.~Xu, ``The meta distribution of sinr in uav-assisted cellular networks,'' \emph{IEEE Trans. Commun.}, vol.~71, no.~2, pp. 1193--1206, Feb. 2023.

\bibitem{wang2022deployment}
L.~Wang, H.~Zhang, S.~Guo, and D.~Yuan, ``Deployment and association of multiple uavs in uav-assisted cellular networks with the knowledge of statistical user position,'' \emph{IEEE Trans. Wireless Commun.}, vol.~21, no.~8, pp. 6553--6567, Aug. 2022.

\bibitem{khawaja2019survey}
W.~Khawaja, I.~Guvenc, D.~W. Matolak, U.-C. Fiebig, and N.~Schneckenburger, ``A survey of air-to-ground propagation channel modeling for unmanned aerial vehicles,'' \emph{IEEE Commun. Surv. Tutorials}, vol.~21, no.~3, pp. 2361--2391, 2019.

\bibitem{chen2022joint}
G.~Chen, X.~B. Zhai, and C.~Li, ``Joint optimization of trajectory and user association via reinforcement learning for uav-aided data collection in wireless networks,'' \emph{IEEE Trans. Wireless Commun.}, vol.~22, no.~5, pp. 3128--3143, May 2023.

\bibitem{lv2025large}
K.~Lv, S.~Huang, Y.~Yao, W.~Jiang, and Z.~Feng, ``Large language model-empowered energy-efficient multi-uav-assisted mec heterogeneous networks,'' \emph{IEEE Trans. Cognit. Commun. Networking}, vol.~12, pp. 5281--5294, 2026.

\bibitem{gao2024marl}
Q.~Gao, R.~Zhong, H.~Shin, and Y.~Liu, ``Marl-based uav trajectory and beamforming optimization for isac system,'' \emph{IEEE Internet Things J.}, vol.~11, no.~24, pp. 40\,492--40\,505, Dec. 2024.

\bibitem{xu2023soft}
F.~Xu, Y.~Ruan, and Y.~Li, ``Soft actor–critic based 3-d deployment and power allocation in cell-free unmanned aerial vehicle networks,'' \emph{IEEE Wireless Commun. Lett.}, vol.~12, no.~10, pp. 1692--1696, Oct. 2023.

\bibitem{2025trajectory}
D.~D. Souza \emph{et~al.}, ``Trajectory optimization in user-centric distributed massive mimo systems enabled by uav swarms,'' \emph{IEEE Trans. Veh. Technol.}, vol.~74, no.~6, pp. 9252--9268, Jun. 2025.

\bibitem{luukkonen2011modelling}
T.~Luukkonen, ``Modelling and control of quadcopter,'' \emph{Independent Res. Project Appl. Math.}, vol.~22, no.~22, pp. 1--24, 2011.

\bibitem{romero2022time}
A.~Romero, R.~Penicka, and D.~Scaramuzza, ``Time-optimal online replanning for agile quadrotor flight,'' \emph{IEEE Rob. Autom. Lett.}, vol.~7, no.~3, pp. 7730--7737, Jul. 2022.

\bibitem{wu2018joint}
Q.~Wu, Y.~Zeng, and R.~Zhang, ``Joint trajectory and communication design for multi-uav enabled wireless networks,'' \emph{IEEE Trans. Wireless Commun.}, vol.~17, no.~3, pp. 2109--2121, Mar. 2018.

\bibitem{wang2025drl}
Y.~Wang, Y.~Hou, J.~Hu, G.~Mu, Q.~Cui, and X.~Tao, ``Drl-based resource allocation and computation offloading in space-air-ground integrated network for iot,'' in \emph{Proc. IEEE 11th World Forum on Internet of Things (WF-IoT)}.\hskip 1em plus 0.5em minus 0.4em\relax IEEE, 2025, pp. 1--6.

\bibitem{li2022applications}
T.~Li \emph{et~al.}, ``Applications of multi-agent reinforcement learning in future internet: A comprehensive survey,'' \emph{IEEE Commun. Surv. Tutorials}, vol.~24, no.~2, pp. 1240--1279, 2022.

\bibitem{zhong2021multi}
R.~Zhong, X.~Liu, Y.~Liu, and Y.~Chen, ``Multi-agent reinforcement learning in noma-aided uav networks for cellular offloading,'' \emph{IEEE Trans. Wireless Commun.}, vol.~21, no.~3, pp. 1498--1512, Mar. 2022.

\bibitem{eschmann2021reward}
J.~Eschmann, ``Reward function design in reinforcement learning,'' in \emph{Reinforcement Learning Algorithms: Analysis and Applications}.\hskip 1em plus 0.5em minus 0.4em\relax Springer, 2021, pp. 25--33.

\bibitem{du2019liir}
Y.~Du, L.~Han, M.~Fang, J.~Liu, T.~Dai, and D.~Tao, ``Liir: Learning individual intrinsic reward in multi-agent reinforcement learning,'' in \emph{Proc. NeurIPS}, vol.~32, 2019.

\bibitem{hadfield2017inverse}
D.~Hadfield-Menell, S.~Milli, P.~Abbeel, S.~J. Russell, and A.~Dragan, ``Inverse reward design,'' in \emph{Proc. NeurIPS}, vol.~30, 2017.

\bibitem{booth2023perils}
S.~Booth, W.~B. Knox, J.~Shah, S.~Niekum, P.~Stone, and A.~Allievi, ``The perils of trial-and-error reward design: misdesign through overfitting and invalid task specifications,'' in \emph{Proc. AAAI}, vol.~37, no.~5, 2023, pp. 5920--5929.

\bibitem{cai2025tutorial}
L.~Cai \emph{et~al.}, ``Tutorial on large language model-enhanced reinforcement learning for wireless networks,'' \emph{arXiv preprint arXiv:2512.03722}, 2025.

\bibitem{zheng2026large}
J.~Zheng \emph{et~al.}, ``Large language model-enabled reinforcement learning for wireless network optimization,'' \emph{IEEE Commun. Mag.}, vol.~64, no.~4, pp. 82--89, Apr. 2026.

\bibitem{cai2025large}
L.~Cai \emph{et~al.}, ``Large language model-enhanced reinforcement learning for low-altitude economy networking,'' \emph{arXiv preprint arXiv:2505.21045}, 2025.

\bibitem{cui2025overview}
Q.~Cui \emph{et~al.}, ``Overview of ai and communication for 6g network: Fundamentals, challenges, and future research opportunities,'' \emph{Sci. China Inf. Sci.}, vol.~68, no.~7, p. 171301, 2025.

\bibitem{kwon2023reward}
M.~Kwon, S.~M. Xie, K.~Bullard, and D.~Sadigh, ``Reward design with language models,'' \emph{arXiv preprint arXiv:2303.00001}, 2023.

\bibitem{xie2024text2reward}
T.~Xie \emph{et~al.}, ``Text2reward: Reward shaping with language models for reinforcement learning,'' in \emph{Proc. ICLR}, 2024, pp. 35\,663--35\,699.

\bibitem{ma2024eureka}
Y.~J. Ma \emph{et~al.}, ``Eureka: Human-level reward design via coding large language models,'' in \emph{Proc. ICLR}, 2024, pp. 26\,516--26\,560.

\bibitem{li2025efficient}
Y.~Li \emph{et~al.}, ``Efficient onboard vision-language inference in uav-enabled low-altitude economy networks via llm-enhanced optimization,'' \emph{arXiv preprint arXiv:2510.10028}, 2025.

\bibitem{wang2024optimal}
Z.~Wang, J.~Zhang, H.~Lei, D.~Niyato, and B.~Ai, ``Optimal bilinear equalizer beamforming design for cell-free massive mimo networks with arbitrary channel estimators,'' \emph{IEEE Trans. Veh. Technol.}, vol.~74, no.~4, pp. 6862--6867, Apr. 2025.

\bibitem{ozdogan2019performance}
{\"O}.~{\"O}zdogan, E.~Bj{\"o}rnson, and J.~Zhang, ``Performance of cell-free massive mimo with rician fading and phase shifts,'' \emph{IEEE Trans. Wireless Commun.}, vol.~18, no.~11, pp. 5299--5315, Nov. 2019.

\bibitem{wang2025optimal}
Z.~Wang, J.~Zhang, E.~Björnson, D.~Niyato, and B.~Ai, ``Optimal bilinear equalizer for cell-free massive mimo systems over correlated rician channels,'' \emph{IEEE Trans. Signal Process.}, vol.~73, pp. 1501--1517, 2025.

\bibitem{bjornson2017massive}
E.~Bj{\"o}rnson, J.~Hoydis, and L.~Sanguinetti, ``Massive mimo networks: Spectral, energy, and hardware efficiency,'' \emph{Found. Trends Signal Process.}, vol.~11, no. 3-4, pp. 154--655, 2017.

\bibitem{bouabdallah2007full}
S.~Bouabdallah and R.~Siegwart, ``Full control of a quadrotor,'' in \emph{Proc. IEEE/RSJ Int. Conf. Intell. Robots Syst. (IROS)}, 2007, pp. 153--158.

\bibitem{evans1991new}
D.~J. Evans, ``A new 4th order runge-kutta method for initial value problems with error control,'' \emph{Int. J. Comput. Math.}, vol.~39, no. 3-4, pp. 217--227, 1991.

\bibitem{hoffmann2007quadrotor}
G.~Hoffmann, H.~Huang, S.~Waslander, and C.~Tomlin, ``Quadrotor helicopter flight dynamics and control: Theory and experiment,'' in \emph{Proc. AIAA Guid., Navigation Control Conf. Exhib.}, 2007, p. 6461.

\bibitem{ward2022development}
S.~Ward and T.~Fields, ``Development and viability of an inverted descent quadrotor for precision aerial delivery,'' in \emph{Proc. AIAA SCITECH 2022 Forum}, 2022, p. 2480.

\bibitem{lowe2017multi}
R.~Lowe, Y.~I. Wu, A.~Tamar, J.~Harb, O.~Pieter~Abbeel, and I.~Mordatch, ``Multi-agent actor-critic for mixed cooperative-competitive environments,'' in \emph{Proc. NeurIPS}, vol.~30, 2017.

\bibitem{liu2022meta}
R.~Liu, F.~Bai, Y.~Du, and Y.~Yang, ``Meta-reward-net: Implicitly differentiable reward learning for preference-based reinforcement learning,'' in \emph{Proc. NeurIPS}, vol.~35, 2022, pp. 22\,270--22\,284.

\bibitem{li2024auto}
H.~Li \emph{et~al.}, ``Auto mc-reward: Automated dense reward design with large language models for minecraft,'' in \emph{Proc. IEEE/CVF Conf. Comput. Vis. Pattern Recognit. (CVPR)}, 2024, pp. 16\,426--16\,435.

\bibitem{wang2023augmenting}
W.~Wang \emph{et~al.}, ``Augmenting language models with long-term memory,'' \emph{Proc. NeurIPS}, vol.~36, pp. 74\,530--74\,543, 2023.

\bibitem{kanungo2002efficient}
T.~Kanungo, D.~Mount, N.~Netanyahu, C.~Piatko, R.~Silverman, and A.~Wu, ``An efficient k-means clustering algorithm: analysis and implementation,'' \emph{IEEE Trans. Pattern Anal. Mach. Intell.}, vol.~24, no.~7, pp. 881--892, Jul. 2002.

\end{thebibliography}

\vfill

\end{document}